# Thermal and Vibrational Properties of Thermoelectric ZnSb – Exploring the Origin of Low Thermal Conductivity

A. Fischer,[1] E.-W. Scheidt,[1] W. Scherer,[1] D. Benson,[2] Y. Wu,[3,†] D. Eklöf,[4] U. Häussermann[4]

[1]*Department of Physics, Augsburg University, D-86159 Augsburg, Germany*
[2]*Department of Physics, Arizona State University, Tempe, Arizona 85287-1504*
[3]*Department of Chemistry and Biochemistry, Arizona State University, Tempe, Arizona 85287-1604*
[4]*Department of Materials and Environmental Chemistry, Stockholm University, S-10691 Stockholm, Sweden*

† present address: *Department of Physics, Tsinghua University, Beijing, China, 100084*

**Abstract**

The intermetallic compound ZnSb is an interesting thermoelectric material, largely due to its low lattice thermal conductivity. The origin of the low thermal conductivity has so far been speculative. Using multi-temperature single crystal X-ray diffraction (9 – 400 K) and powder X-ray diffraction (300 – 725 K) measurements we characterized the volume expansion and the evolution of structural properties with temperature and identify an increasingly anharmonic behavior of the Zn atoms. From a combination of Raman spectroscopy and first principles calculations of phonons we consolidate the presence of low-energy optic modes with wavenumbers below 60 cm$^{-1}$. Heat capacity measurements between 2 and 400 K can be well described by a Debye-Einstein model containing one Debye and two Einstein contributions with temperatures $\Theta_D = 197$K, $\Theta_{E1} = 77$ K and $\Theta_{E2} = 271$ K as well as a significant contribution due to anharmonicity above 150 K. The presence of a multitude of weakly dispersed low-energy optical modes (which couple with the acoustic, heat carrying phonons) combined with anharmonic thermal behavior provides an effective mechanism for low lattice thermal conductivity. The peculiar vibrational properties of ZnSb are attributed to its chemical bonding properties which are characterized by multicenter bonded structural entities. We argue that the proposed mechanism to explain the low lattice thermal conductivity of ZnSb might also control the thermoelectric properties of electron poor semiconductors, such as $Zn_4Sb_3$, $CdSb$, $Cd_4Sb_3$, $Cd_{13-x}In_yZn_{10}$, and $Zn_5Sb_4In_{2-\delta}$.

# I. Introduction

The binary Zn-Sb system affords two semiconductor phases, ZnSb and β-$Zn_4Sb_3$, which have been known as thermoelectric materials since the 1960s [1,2]. Especially β-$Zn_4Sb_3$ shows an excellent thermoelectric performance in the temperature range of 450 – 650 K and has been intensively studied during the past 15 years [3,4]. This material is distinguished by an extraordinarily low lattice thermal conductivity in the range of 0.5 – 0.6 W/mK at RT which is characteristic for vitreous materials [5]. Accordingly, this feature is truly remarkable for a crystalline binary compound. Recently it has been shown that also ZnSb has an inherently low thermal conductivity [6] and that doping with Ag/Cu (leading to $Cu_3Sb$ and $Ag_3Sb$ nanoparticle inclusions) can produce materials with thermoelectric figure of merits almost in par with β-$Zn_4Sb_3$ [7,8].

The origin of the low lattice thermal conductivity of β-$Zn_4Sb_3$ and ZnSb is puzzling. Constituting elements are not particularly heavy. In 2004 a careful structural analysis of β-$Zn_4Sb_3$ revealed intricately disordered Zn atoms. Naturally, this structural disorder was then associated with low thermal conductivity [9]. However, the discovery of $Cd_{13-x}In_yZn_{10}$ (x ≈ 2.7, y ≈ 1.5), which crystallizes in a disorder-free variant of the β-$Zn_4Sb_3$ structure and shows a comparable low thermal conductivity, casted doubts into the significance of structural disorder to low thermal conductivity [10]. Recently, peculiarities in the dynamic behavior − notably Einstein rattling of dumbbell units build from Sb atoms [11] and/or loosely bonded Zn atoms as evidenced from the superionic behavior of β-$Zn_4Sb_3$ [12,13] have been put forward as possible reasons. It is not clear if these aspects would also apply to ZnSb. Furthermore, it should be noted that some ternary derivatives of zinc antimonides, like $Zn_5Sb_4In_{2-\delta}$ (δ ≈ 0.15), display similar low thermal conductivities as the binaries [14,15].

Zinc antimony compounds and their derivatives clearly have diverse crystal structures with features that individually could explain low thermal conductivity (e.g. presence of various forms of disorder, large sized unit cells). Yet, common to all systems is a covalent framework structure containing multicenter bonded structural entities. It is therefore tempting to propose such an electronic motif as a common, underlying characteristics that controls the unusual inherent low lattice thermal conductivity for all these systems. Indeed, peculiarities in the dynamic behavior could originate in the bonding properties of the framework structures, that is, multicenter bonded atoms may cause the incidence of pronounced anharmonic vibrational motion and/or give rise to localized low energy optical modes (which couple with the acoustic, heat carrying phonons). The presence of localized low energy optic modes as a consequence of multicenter bonded structural entities would provide a sound physical basis for the empirical concept of "dumbbell" rattling recently pursued for $Zn_4Sb_3$ [11]. This hypothesis has been put forward earlier [16], and recent theoretical investigations by Jund *et al.* [17] and Bjerg *et al.* [18] seem to confirm it.

The aim of the present work is to shed more light on this issue. From heat capacity measurements we reveal a great similarity of the vibrational properties of ZnSb and β-$Zn_4Sb_3$. From the analysis of the temperature dependence of the structural and atomic displacement parameters we find that Zn atoms display an increasingly anharmonic thermal motion at temperatures above 200 K. From phonon dispersion calculations and Raman spectroscopy we consolidate the presence of low energy optic modes in ZnSb as a consequence of multicenter bonded rhomboid ring entities $Zn_2Sb_2$. Those rings are also present in the structure of β-$Zn_4Sb_3$.

## II. Methods

### A. Synthesis

Bulk samples of polycrystalline ZnSb were prepared from mixtures of zinc granules (ACBR, 99.99%) and antimony shots (ABCR, 99.999%) using a slight excess (2%) of Zn. Batches with a total mass between 0.5 and 1 g were loaded in fused silica tubes which were flame sealed in a dynamic vacuum (< $10^{-5}$ bar). The mixture of elements was melted with a torch while shaking the ampoule vigorously, and then quenched in water. The obtained ingot was ground to a powder which was subsequently sealed in a fused silica tube and annealed for 5 days at 783 K. So obtained polycrystalline ZnSb was phase pure according to powder X-ray diffraction analysis. Bulk sample was partly consolidated by spark plasma sintering (SPS) at 450 °C and 75 MPa for 5 min using a Dr. Sinter SCM 5000 instrument. In the following we refer to this sample as "SPS sample". Crystal specimens of ZnSb with sizes of several mm were prepared from reaction mixtures with 23 at.% Zn, 27 at.% Sb and 50 at.% Bi, where Bi serves as a flux medium. Starting materials were Zn (granules, 99.99%), Sb (powder, 99.5%) and Bi (pieces, 99.999%). The procedure of crystal growth and separation is described in ref. [6].

### B. Powder X-ray diffraction

For powder X-ray diffraction (PXRD) analysis polycrystalline ZnSb bulk sample was finely ground. High temperature PXRD studies were performed on a Panalytical X'Pert PRO instrument with Cu Kα radiation in θ-2θ diffraction geometry. The powder sample was heated in dynamic vacuum (~ $10^{-5}$ bar ) to 723 K using an Anton Paar XRK 900 chamber equipped with Be windows and connected to a temperature controller. Data in a 2θ range 10 – 60° were collected from 323 K in steps of 50 K, 100 min acquisition time and 10°/min heating rate between the steps. Data collection was initiated when the temperature had stabilized. A room temperature diffraction pattern was collected on a Panalytical X'Pert Alpha1 diffractometer operated with Cu K$α_1$ radiation where the sample was mounted on a Si wafer zero-background holder. Rietveld refinement of PXRD data was performed using the Jana2006 package [19]. The details of the refinement results are given in the supporting information.

### C. Single Crystal X-Ray diffraction

For single crystal X-ray diffraction (SCXRD) analysis, crystals were selected among specimens obtained from flux synthesis. High resolution data were collected using a BRUKER SMART-APEX diffractometer equipped with a D8 goniometer and an INCOATEC IμS Ag micro source ($\lambda$ = 0.56087 Å) employing Helios mirror optics. Measurements were performed at several temperatures between 80 K and 400 K with an Oxford cryostream cooling unit. The frames were integrated with the Bruker SAINT software package [20] using a narrow-frame algorithm. A numerical, face-indexed absorption correction was applied using SADABS [21]. Data at 9 K were collected on a MAR345 imaging plate detector system with a rotating anode generator (Bruker FR591, $\lambda$ = 0.71073 Å) using a a displex cryo-system and a Huber 512.1 Eulerian cradle. The frames were integrated using the EVAL14 software and a numerical absorption correction and inter-frame scaling correction was performed using SADABS. Face indexing and crystal shape determination were however performed using the SMART-APEX diffractometer. The EUHEDRAL software was used to determine the crystal orientation during the measurements in the displex cryo-system for the subsequent numerical absorption corrections. All structure refinements were performed with the program package JANA2006 [19]. All datasets were corrected for extinction (type 2) during the refinements. In order to exclude a possible bias between the correction of absorption and extinction effects, which mainly affect the data at low resolution, atomic displacement parameter (ADP) values were also derived from refinements of high order Bragg reflection ($\sin\Theta/\lambda > 0.6$ Å$^{-3}$). However, the ADP values derived independently from both methods did not differ significantly. The details of refinement summary and parameters are given in the supporting information.

### D. Raman spectroscopy

Raman spectroscopy investigations were performed on ZnSb crystal specimens obtained from flux growth and on a disk-shaped specimen with 12 mm diameter obtained from SPS consolidation. For comparison also elemental Sb (as purchased) was investigated. Raman spectra were measured using a Labram HR 800 spectrometer. The instrument is equipped with an 800 mm focal length spectrograph and an air cooled (-70 ºC), back thinned CCD detector. Samples were excited using an air cooled double frequency Nd:YAG laser (532 nm) with a reduced input laser power of 0.56 mW and an air cooled intra cavity regulated laser diode (785nm) with reduced laser power of 0.88 mW. Raman spectra were collected with an exposure time of 60 s, accumulation number of 10, and using a 1800 grooves/mm grating.

### E. Heat capacity measurements

For heat capacity measurement a crystal specimen (14.8 mg), and a SPS sintered piece (28.0 mg) were used. The heat capacity was measured between 2 K and 290 K (49 points distributed logarithmically) and between 330 K and 400 K in 3 K increments using a quasi-adiabatic step heating technique as implemented in the Physical Property Measurement System (PPMS) by Quantum Design. The samples were thermally connected to the platform of the sample-holder via small amount of Apiezon-N grease for region $T < 300K$ and Apiezon-H for $T > 300K$ (both typically $0.1 - 0.3$ mg). The uncertainty for this measurement technique is estimated to be lower than 5%.

### F. Computation

The phonon dispersion relations and thermodynamic functions of ZnSb were calculated via the Abinit program package [22–24] and employing the generalized gradient approximation (GGA) with the Perdew-Burke-Ernzerhof (PBE) parameterization [25,26]. GGA-PBE pseudopotentials were provided by the Abinit website. These pseudopotentials are norm conserving and were generated using the fhi98PP package [27]. A $6 \times 6 \times 6$ Monkhorst

Pack [28] k-point grid was used for electronic integration and a 2 × 2 × 2 q-grid was used for calculations of the dynamical matrix elements. A plane wave energy cutoff of 35 Hartree (~950 eV) was employed. Prior to the phonon dispersion calculations ZnSb was relaxed with respect to lattice parameters and atomic positions with forces converged to better than $1 \times 10^{-3}$ eV/Å. From the phonon density of states (PDOS) it is possible to obtain the thermodynamic functions of a material [29]. Within the harmonic approximation the contribution from the phonons per unit cell to the internal energy $\Delta E$, the Helmholtz free energy $\Delta F = \Delta E - TS$, the constant volume specific heat $C_V$, and the entropy $S$, at temperature $T$ are given by:

$$\Delta F = 3nk_B T \int_0^{\omega_L} \ln\left\{2\sinh\frac{\hbar\omega}{2k_B T}\right\} g(\omega) d\omega \qquad (1)$$

$$\Delta E = 3n\frac{\hbar}{2} \int_0^{\omega_L} \omega \coth\left(\frac{\hbar\omega}{2k_B T}\right) g(\omega) d\omega \qquad (2)$$

$$C_v = 3nk_B \int_0^{\omega_L} \left(\frac{\hbar\omega}{2k_B T}\right)^2 \operatorname{csch}^2\left(\frac{\hbar\omega}{2k_B T}\right) g(\omega) d\omega \qquad (3)$$

$$S = 3nk_B \int_0^{\omega_L} \left[\frac{\hbar\omega}{2k_B T} \coth\frac{\hbar\omega}{2k_B T} - \ln\left\{2\sinh\frac{\hbar\omega}{2k_B T}\right\}\right] g(\omega) d\omega, \qquad (4)$$

where $k_B$ is the Boltzmann constant, $\omega$ is the phonon frequency, $\omega_L$ is the largest frequency and $n$ is the number of atoms in the unit cell and $g(\omega)$ is the PDOS and csch is the hyperbolic cosecant.

Atomic displacement parameters (ADPs) can be obtained from the PDOS with the knowledge of the corresponding eigenvectors $e_{n\alpha}^l(\vec{q})$ and the partial density of states of the phonons. The partial density of state of the phonons is given by [30]:

$$g_{n\alpha\beta}(\omega) = \sum_{\vec{q},l} \delta(\omega - \omega_l(\vec{q}))\left\langle e_{n\alpha}^{l*}(\vec{q}) e_{n\beta}^l(\vec{q})\right\rangle, \qquad (5)$$

where $\alpha$ and $\beta$ represent the $x$, $y$ or $z$ Cartesian coordinates, $n$ is the atom index, $l$ is phonon mode and $\vec{q}$ is the wavevector in the Brillouin zone. The ADPs $U_{nij}$ are calculated as:

$$B_{n\alpha\beta} = \int_0^\infty \frac{8\pi^2 \hbar}{M_n \omega} g_{n\alpha\beta}(\omega)\left(\frac{1}{\exp\left(\frac{\hbar\omega}{k_B T}\right) - 1} + \frac{1}{2}\right) d\omega, \qquad (6)$$

with

$$U_{nij} = \frac{1}{8\pi^2} \sum \frac{b_{i\alpha} B_{n\alpha\beta} b_{j\beta}}{|\vec{b}_i||\vec{b}_j|}, \qquad (7)$$

where $M_n$ is the mass of atom $n$ and $\vec{b}_i$ ($i, j = 1, 2$ or 3) are the crystal reciprocal lattice vectors. $U_{nij}$ is a tensor that defines the average motions of the atoms within the compounds (with dimensions of length squared). The isotropic ADPs $U_{iso}^n$ can be derived directly from the anisotropic $U_{nij}$ values by:

$$U_{iso}^n = \frac{1}{3}\sum_i U_{nii}. \qquad (8)$$

### III. Results and Discussion

#### A. Temperature dependent structural properties and thermal stability

ZnSb crystallizes with an orthorhombic *Pbca* structure (the CdSb type) which contains 8 formula units in the unit cell [31]. Both kinds of atoms are situated on the general position 8c. The structure may be described as being built from rhomboid rings $Zn_2Sb_2$ which are arranged in layers and linked to 10 neighboring rings. Fig. 1a shows the arrangement of layers along the [001] direction. (Note, that because of the three axial glides this description holds for any direction). In a layer each ring is surrounded by six neighboring ones; two are attached to each Sb atom and one to each Zn atom of one $Zn_2Sb_2$ moiety. Additionally each atom will then bind to a ring in an adjacent layer. As a result Zn and Sb atoms in the ZnSb structure attain a peculiar five-fold coordination by one like and four unlike neighbors (Fig. 1b). A ring and its linkage to neighboring ones is shown in Fig. 1c.

According to refs. [32] and [33] the rhomboid ring represents a four center, four electron (4c4e) bonded entity which is connected via 2c2e bonds to the 10 neighboring ones. This bonding model provides an electron precise situation for ZnSb. There are six distinct nearest neighbor distances in the ZnSb structure which are all captured within the rhomboid ring and its connectivity (cf. Fig 1c). The different bonding motifs (rhomboid ring multi-center 4c4e and connecting 2c2e bonds) can be recognized in the distribution of interatomic distances: Zn-Sb interatomic distances within a ring (r-type distances) are about 0.1 Å larger than the ring connecting ones (c-type distances). The short Zn-Zn distance (2.78 Å) is part of the multi-center bonding motif while the short Sb-Sb distance of around 2.82 Å corresponds to a ring-linking 2e2c bond. These nearest neighbor distances associated with bonding interactions are well separated from the next nearest ones, starting off above 3.5 Å.

The rhomboid ring $Zn_2Sb_2$ motif is also the central feature of the $Zn_4Sb_3$ structure (Fig. 1d) [16,33]. Here rings are condensed into chains by sharing common Sb atoms (Fig. 1e). These Sb atoms (termed Sb1) then attain a six-coordination by Zn atoms by connecting with two neighboring chains via 2c2e bonds. The linkage of chains in the final framework is completed by additional Sb atoms (termed Sb2) forming dumbbells (Fig. 1e). Each dumbbell unit has six chain-linking Sb2-Zn contacts, also corresponding to 2c2e bonds while the Sb2 atoms are tetrahedrally coordinated. From inelastic neutron scattering measurements Schweika et al. concluded that vibrations of these Sb2 dumbbells against their Zn coordination environment correspond to a soft, localized and strongly anharmonic mode [11]. Consequently the authors suggested that this "dumbbell rattling" rather than structural disorder is the origin of low thermal conductivity of β-Zn$_4$Sb$_3$. We emphasize that "dumbbells" of Sb atoms also occur in the ZnSb structure, however, in ZnSb they correspond to direct links of neighboring rhomboid rings and their coordination environment is rather different from the one in β-Zn$_4$Sb$_3$.

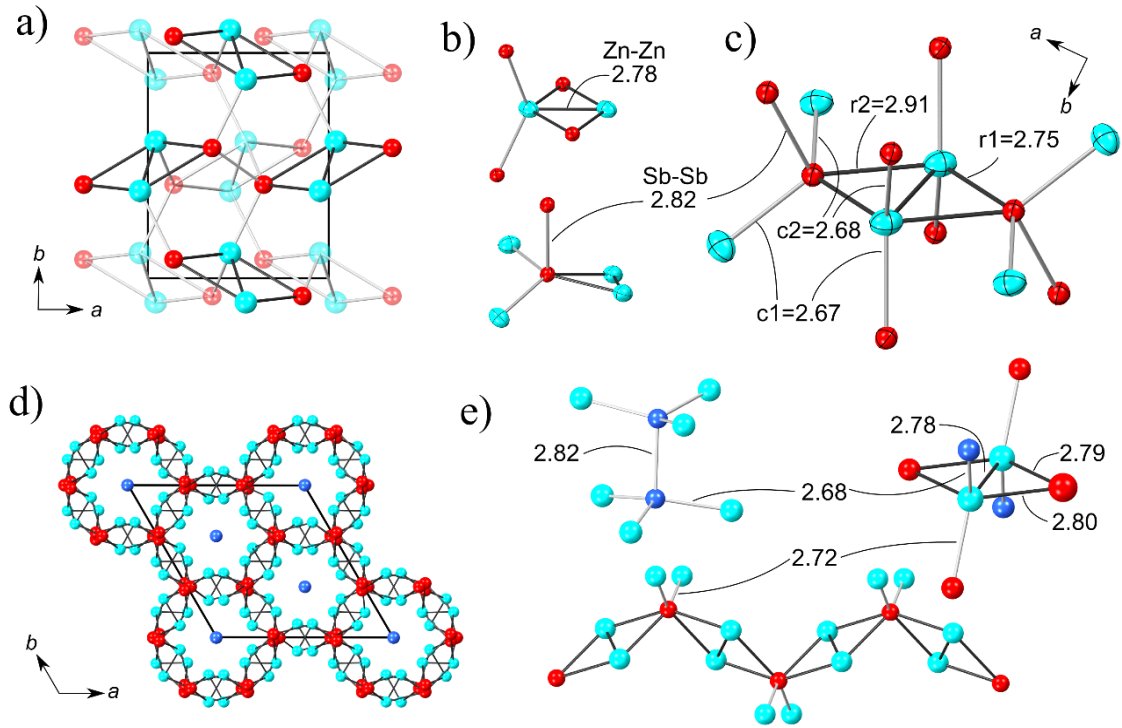

Figure 1: (a) The orthorhombic crystal structure of ZnSb built from layers (dark and pale colors) of rhomboid rings Zn$_2$Sb$_2$. Cyan and red balls denote Zn and Sb atoms, respectively (b) Local five-coordination of Zn and Sb atoms. The inserted numbers indicate interatomic distances in Å at room temperature. (c) Rhomboid ring Zn$_2$Sb$_2$ and its connectivity to neighboring rings. Distances are distinguished as c- (connecting) and r- (ring) type. The center of a ring corresponds to a center of inversion. The thermal ellipsoids correspond to 90% probability density at room temperature obtained from single crystal XRD measurements. (d) Idealized Zn$_6$Sb$_5$ framework of rhombohedral β-Zn$_4$Sb$_3$ along the [001] direction. The framework is built from chains of condensed rhomboid rings Zn$_2$Sb1$_{2/2}$ (cf. Fig. 1e) and consists of channels which are stuffed by Sb2 atoms (depicted as blue balls). Sb2-Zn bonds are omitted for clarity. (e) Structural fragments and local coordination of Zn, Sb1, and Sb2 atoms in β-Zn$_4$Sb$_3$.

To investigate the effect of temperature to the structural parameters of ZnSb we performed single crystal X-ray diffraction (SCXRD) measurements from 9 to 400 K, and powder X-ray diffraction (PXRD) measurements from room temperature up to 723 K. Figs. 2a and 2b show lattice parameters and unit cell volume as a function of temperature, respectively. Lattice parameters refined from PXRD data are more accurate. They increase linearly in the range 300 – 675 K. The room temperature volume expansion coefficient $α_V$(300 K) is estimated as 4.15 × 10$^{-5}$ K$^{-1}$ which is comparable to many metals (e.g. Cu: 5.1 × 10$^{-5}$ K$^{-1}$, Au: 4.2 × 10$^{-5}$ K$^{-1}$). We note a noticeable anisotropy of the room temperature linear thermal expansion coefficients for the individual lattice parameters: $\alpha_c = 0.404 \times 10^{-5}$ K$^{-1}$, $\alpha_b = 1.25 \times 10^{-5}$ K and $\alpha_a = 2.45 \times 10^{-5}$ K. The ratio $\alpha_c : \alpha_b : \alpha_a$ corresponds roughly to 1:3:6. The variation of interatomic distances obtained from the refinement of the SCXRD data is shown in Fig. 2c in the range of 9 – 400 K. Two of the six nearest neighbor distances show pronounced temperature variation: the intra ring distances r2 and the Zn-Zn separation increase roughly linearly from 2.87 to 2.93 Å and from 2.74 to 2.80 Å with increasing temperature from 9 to 400K, respectively. The remaining distances increase in the same temperature range only by 0.01 to 0.015 Å, corresponding to a change of 0.4 – 0.5%.

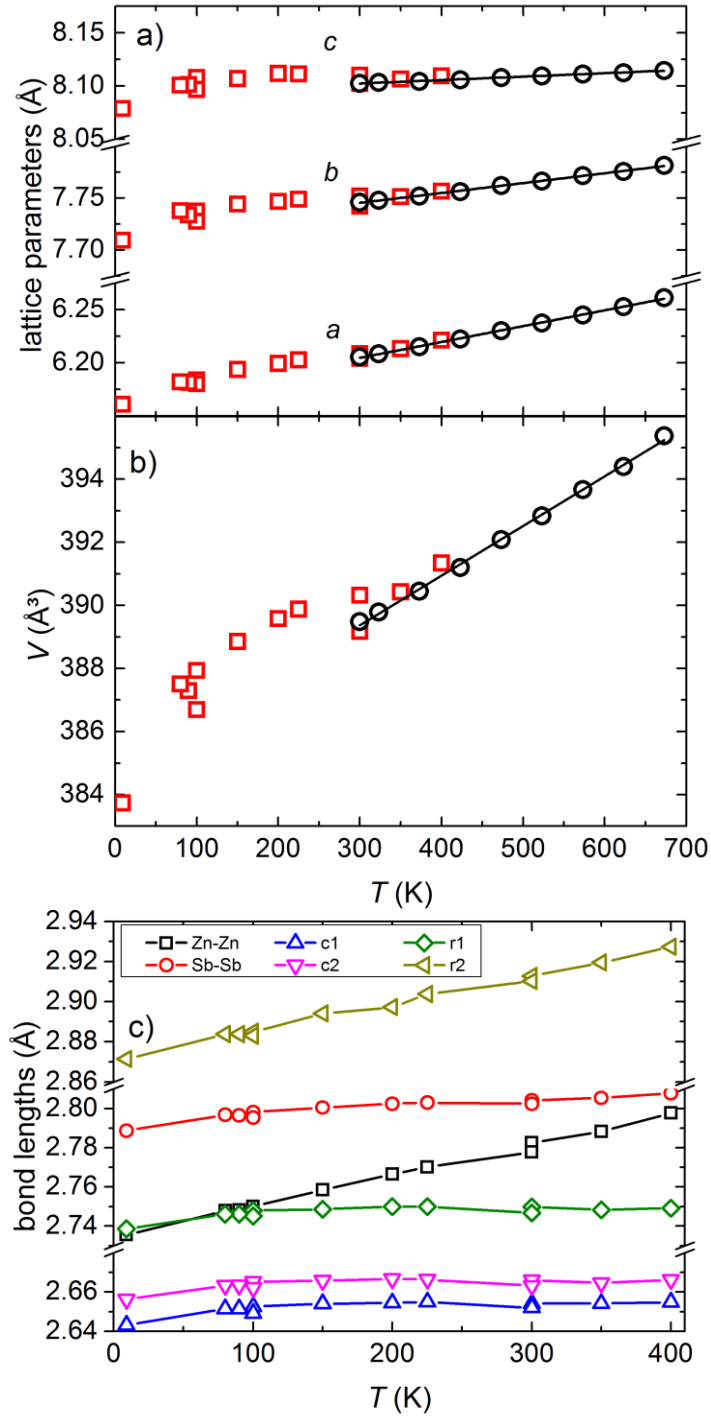

Figure 2. (a) Lattice parameters and (b) unit cell volume of ZnSb as a function of temperature.

Squares and circles denote parameters obtained from SCXRD and PXRD refinements, respectively. The solid line represents a linear fit to the PXRD data. (c) Variation of interatomic distances in ZnSb as a function of temperature. Data were obtained from refinements of SCXRD data including anharmonic atomic displacement parameters (using a Gram-Charlier expansion). For the assignment of the individual bond lengths, see Fig. 1. The size of the symbols used in (a) – (c) exceeds by far the standard deviation of the corresponding data point.

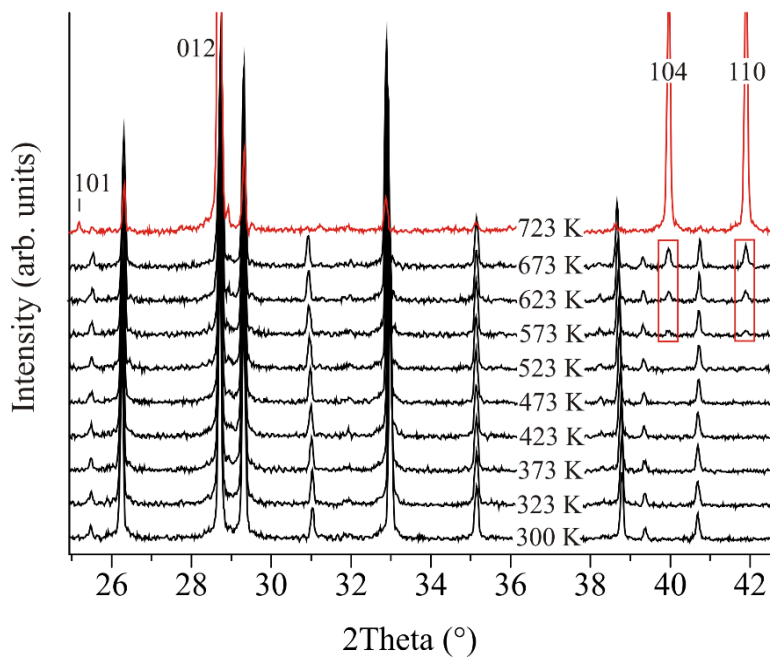

Figure 3. PXRD patterns (Cu K$\alpha$ radiation) of ZnSb at temperatures between 300 and 723 K in a 2$\Theta$ range of 25 – 42.5°. At 573 K reflections from elemental antimony occur (marked with boxes). At 723 K the sample decomposition is already severe and a mixture of ZnSb and elemental Sb (*R*-3*m*) with a molar ratio of approx. 1:6 is observed. Bragg reflections of the Sb impurity phase are marked by their corresponding Miller indices.

Fig. 3 shows the PXRD patterns of ZnSb upon heating between 300 and 723 K in a dynamic vacuum. At 573 K the onset of the sample decomposition is signaled by the occurrence of reflections of elemental antimony. Phase analyses by Rietveld refinements yield an amount of elemental Sb of 1.4(2) wt%. At 723 K (and about 5 h after the 573 K measurement) the ZnSb sample is almost completely decomposed. The Sb phase fraction increased to 78.4(4) wt%. The presence of dynamic vacuum conditions clearly promotes the decomposition of ZnSb (into elemental Sb and Zn vapor) which otherwise, according to the Zn-Sb phase diagram [34], represents a thermodynamically stable phase up to at least 800 K. The same phenomenon has been observed for β-$Zn_4Sb_3$: Heating $Zn_4Sb_3$ in a dynamic vacuum leads to the formation of ZnSb at 623 K which in turn decomposes above 693 K [35]. Thus the thermal stability of ZnSb and β-$Zn_4Sb_3$ under dynamic vacuum conditions is comparably low.

## B. Heat capacity measurements

Heat capacity studies of ZnSb have been performed already in the 1970s by Mamedova *et al.* and Danilenko *et al.* [36,37]. The key results of these studies are compared to our studies in Fig. 4. The measurement of Danilenko et al. exceeds the Dulong-Petit limit of $C_v = 3R$ per mole and atom (49.9 J/mol-K for ZnSb) at temperatures above 240 K. A similar behavior is also displayed by our SPS sample, whereas our crystalline sample reaches this limit just above 330 K, in better agreement with the results of Mamedova *et al.*, (especially below 120 K) whose ZnSb sample actually contained 3 wt.% CdSb [36]. The higher $C_p$ values for the SPS sample may be explained by defects or stress/strain introduced by the SPS consolidation process.

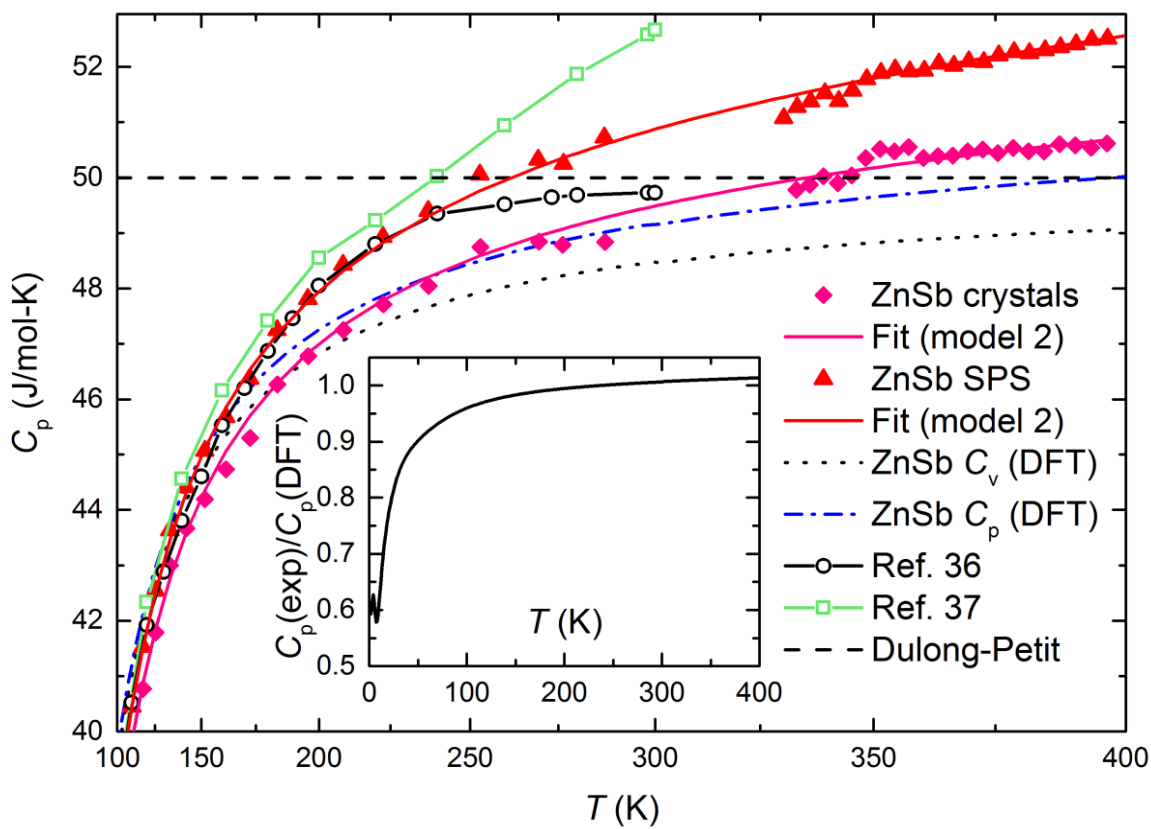

Figure 4: Temperature dependent heat capacity $C_p(T)$ (per formula unit) of crystalline and SPS sintered ZnSb samples (pink diamonds and red triangles, respectively) and their corresponding fits (model 2) as pink and red solid lines, respectively. The Dulong-Petit limit of $3R$ is marked as a black broken horizontal line. Earlier measurements of Danilenko *et al.* [37] and Mamedova *et al.* [36] have been added for comparison (black circles and green squares, respectively, interconnection lines as a guide for the eye). DFT calculated heat capacities $C_v$ and $C_p$ of ZnSb are shown as dashed, black and dash-dotted, blue lines, respectively. The theoretical $C_p$ values have been estimated via the thermal expansion coefficient from the PXRD measurements (see below). The inset shows the ratio of experimental and theoretical $C_p$ values.

As a next step we attempted to describe the $C_p$ data by a simple Debye-Einstein model, which assumes a Debye-type behavior for the three acoustic phonon branches and an Einstein (independent oscillator) behavior for the contributions from the optical branches [38]. Additionally, a linear expansion coefficient $A_1$ is employed to account for the volume dependence of $C_p$ in the high temperature range. The model (in the following referred as "model 1") used can be expressed by the equation:

$$C_v(T) = 3R\left(D(T, \Theta_D) + \sum_{i=1}^{k} c_i\, E_i(T, \Theta_{E,i})\right) \qquad C_v(\infty) = 3nR \qquad (9)$$

where $D$ is the temperature dependent Debye contribution, $E_i$ the temperature dependent Einstein contributions, $c_i$ their respective coefficients and $n$ the number of atoms per formula unit. The quasi-harmonic approximation of $C_p$ can then be described as

$$C_p(T) = C_v(T)\bigl(1 + \tilde{C}_v(T)\, A_1 T\bigr) \qquad \tilde{C}_v(T) = \frac{C_v(T)}{C_v(\infty)} \qquad (10)$$

The Debye contribution was fixed to $3R$, whereas the contributions of the Einstein terms were refined individually but constrained to a sum of $3R$ in order to fulfill the Dulong-Petit law. This simplistic model (model 1) using 1 Debye and 2 Einstein components ($E_1 + E_2$) produced a highly satisfactory fit (Fig. 4 and Fig. 5a), apart from the region at very low temperatures (below 10 K, which is also characterized by non-constant $C/T^3$ values (Fig. 5b)).

The latter observation indicates deviations from a pure Debye behavior at low temperatures (only $\omega^2$-dependence of the PDOS; $g(\omega) \sim \omega^2$ in Equation 3). Additional quartic ($g(\omega) \sim \omega^4$) contributions are, however, frequently observed in binary II-VI or III-V semiconductors [39]. We therefore included such a Non-Debye (ND) component (model 2, Eq. 11) in our specific heat capacity model of ZnSb. For stability reasons $\Theta_{ND}$ was set in the fitting procedure to the lowest Einstein temperature $\Theta_{E1}$ as suggested by others [40] and the sum of the Debye and Non-Debye components was constrained to $3R$.

$$C_v(T) = 3R\left(c_D D(T, \Theta_D) + c_{ND} ND(T, \Theta_{ND}) + \sum_{i=1}^{k} c_i\, E_i(T, \Theta_{E,i})\right); \quad c_D = (1 - c_{ND}) \qquad (11)$$

Figure 5. (a) Heat capacity $C_p/T$ (per formula unit) of ZnSb (black diamonds) including Debye-Einstein fit (model 2, black solid line) and $Zn_4Sb_3$ (red circles) including the original fit of Schweika *et.-al.* [11] (red, solid line) as well as the improved fit using model 1 (green solid line). The DFT calculated data of ZnSb is shown as a blue, dotted line. (b) $C_p/T^3$ (per formula unit) representation of (a) with additional fit for ZnSb (model 1, black, dashed line) for comparison. The *y*-axis range is scaled according to the number of atoms per formula unit in ZnSb and $Zn_4Sb_3$ by the ratio 2:7, for both $C_p/T$ and $C_p/T^3$.

The fits are detailed in Figs. 5a and 5b as $C_p/T$ and $C_p/T^3$ plots, respectively. The more sophisticated model 2 (black solid line) improves the fit of the $C_p$ for ZnSb in the low temperature regime of 3 – 10 K significantly in comparison with model 1 (dashed black line, c.f. Fig 5b). This stresses the necessity to consider a non-Debye component for a precise description of the temperature dependence of $C/T^3$ in the temperature regime. The Debye temperature $\Theta_D(0)$ of ZnSb as obtained as 248K from the slope of $C/T$ vs. $T^2$ representation for $T < 3K$. Jund *et al.* calculated the Debye temperature of ZnSb from averaged sound velocity data measured by Balazyk *et al.* and obtained a value of 253 K, which is in close agreement with our findings [17,41]. The fitted Einstein temperatures are $\Theta_{E1} = 77.6$ K (54 cm$^{-1}$, 6.7 meV) and $\Theta_{E2} = 276.9$ K (192 cm$^{-1}$, 23.9 meV) refer to the single crystal sample. The results of the fit of the SPS sample data with equation (9 and 10) are very similar. The according coefficients are listed in Table 1.

The temperature dependence of the heat capacity of ZnSb and $Zn_4Sb_3$ is highly related. This is shown in Fig. 5a where our measurement for single crystalline ZnSb is compared with the heat capacity of polycrystalline $Zn_4Sb_3$ as obtained by Schweika *et al.* [11]. The authors described the heat capacity of $Zn_4Sb_3$ by a model containing only one Debye and one Einstein contribution with a ratio of 85:15. The energy of the Einstein mode was very low (5.4 meV, 62 K) and associated with a physical oscillator in the $Zn_4Sb_3$ structure, i.e. the before mentioned rattling of $Sb_2$ dumbbells. However, the Schweika *et al.* model is physically implausible because the majority of optical modes would be assigned a Debye behavior. As a matter of fact, this simple model displays significant discrepancies to the experimental data, especially in the low temperature regime (T < 125 K), as shown in Fig. 5a and Fig 5b. To obtain a proper fit using a Debye-Einstein model similar to ZnSb (model 1), 1 Debye and 3 Einstein terms ($E_1$, $E_2$, $E_3$) are needed (cf. Figs. 5a and 5b). The coefficients are presented in Table 1. As for ZnSb, the Debye contribution was fixed to $3R$, whereas the contributions of the Einstein terms were refined individually. Due to the presence of a phase transition in $Zn_4Sb_3$, data around 250 K were excluded. The necessity of Non-Debye contributions was not justified for $Zn_4Sb_3$.

Table 1: Coefficients obtained from a fits of the experimental heat capacity of ZnSb and $Zn_4Sb_3$ using the models described above.

|  | ZnSb-SC | ZnSb-SC | ZnSb-SPS | ZnSb-calc | $Zn_4Sb_3$* | $Zn_4Sb_3$ ** |
|---|---|---|---|---|---|---|
| model | 1 | 2 | 2 | 2 | 1 | "Schweika" |
| $\Theta_D(0)$ [K] | 248 | 248 | 253 | 210 | 251 | 240 |
| $\Theta_D$ [K] | 197.3 | 195.2 | 200.5 | 167.0 | 137.9 | 251 |
| $c_D$ [atoms] | 1 | 0.90 | 0.93 | 0.95 | 1 | 5.95 |

| $\Theta_{ND}$ [K] | - | 77.6[+] | 77.0[+] | 64.5[+] | - | - |
|---|---|---|---|---|---|---|
| $c_{ND}$ [atoms] | - | 0.10 | 0.07 | 0.05 | - | - |
| $\Theta_{E1}$ [K] | 71.5 | 77.6[+] | 77.0[+] | 64.5[+] | 56.2 | 62 |
| $c_{E1}$ [atoms] | 0.36 | 0.30 | 0.34 | 0.29 | 0.38 | 1.05 |
| $\Theta_{E2}$ [K] | 285.7 | 276.9 | 271.8 | 257.0 | 221.0 | - |
| $c_{E2}$ [atoms] | 0.64 | 0.70 | 0.66 | 0.71 | 4.01 | - |
| $\Theta_{E3}$ [K] | - | - | - | - | 85.4 | - |
| $c_{E3}$ [atoms] | - | - | - | - | 1.60 | - |
| $A_1$ [$10^{-6}K^{-1}$] | 95.2 | 93.6 | 187.4 | 49.9 | 76.7 | - |

\* data from ref. [11]   \*\* original fitting model (three parameter, two component fit) by Schweika et al. [11]   [+] constrained

The fitted Einstein temperatures for $Zn_4Sb_3$ are $\Theta_{E1}$ = 56.2 K (39 cm$^{-1}$, 4.8 meV), $\Theta_{E2}$ = 221.0 K (153 cm$^{-1}$, 19.1 meV) and $\Theta_{E3}$ = 85.4 K (59 cm$^{-1}$, 7.4 meV). Accordingly, our model (model 1) uses two low-energy Einstein modes as compared to one in the Schweika model. The ratio of low- and high-energy Einstein mode contributions ($c_{E1}$+ $c_{E3}$):$c_{E2}$ is about 1:2 which is similar to the ratio $c_{E1}$:$c_{E2}$ for ZnSb. Also the Debye temperature $\Theta_D(0)$ at low temperatures of both compounds is very similar (around 250 K). We conclude that although the fitting equations for the heat capacity of ZnSb and $Zn_4Sb_3$ were somewhat different, their overall temperature dependence is very similar. This indicates that both compounds are characterized by rather similar lattice dynamical and thermal properties.

## C. Lattice dynamics and vibrational properties

Fig. 6a shows phonon dispersion curves obtained from first principles calculations and the corresponding PDOS of ZnSb. There is good agreement between our linear response-based calculations and the calculations by Jund *et al.* and Bjerg *et al.* employing a methodology using a super cells approach [17,18]. The prominent feature of the PDOS is the occurrence of a gap between 125 and 140 cm$^{-1}$. This gap separates 20 modes of high energies from the remaining ones. The high energy modes appear to be split into a 4 + 4 (in the range 170 – 190 cm$^{-1}$) and 8 + 4 pattern (in the range 140 – 170 cm$^{-1}$). We note, that the magnitude of atomic displacements of the Zn and Sb atoms with regard to the individual normal modes of vibration is rather similar – despite the large mass difference and the rather different chemical coordination environments displayed by both atom types (see above). However, some exceptions are observed: Weakly dispersed modes at around 165 cm$^{-1}$ and 145 cm$^{-1}$ are characterized by large displacements of the Sb and Zn atoms, respectively. The mode at 156 cm$^{-1}$ could be approximately described as a stretching mode of $Sb_2$ moieties and is fundamentally different from the dumbbell rattling picture which has been propagated in case of the $Sb_2$ clusters in $Zn_4Sb_3$ [11]. Among the low energy modes (below the gap) the ones around 83 cm$^{-1}$ are distinguished because of their weak dispersion and their characteristic and predominant Zn displacements. Below 30 cm$^{-1}$ the main contribution to the PDOS originates from Sb atoms while the contribution from Zn atoms is negligible. Lastly we note an accumulation of low-energy optic modes with small dispersion in the range between 35 and 60 cm$^{-1}$, which conforms with the earlier calculations of Jund *et* *al.* and Bjerg *et al.* [17,18].

Fig. 6b shows the model PDOS reconstructed from the experimental specific heat data of ZnSb using model 2 (cf. Table 1). Both Einstein temperatures are well reflected in the theoretical PDOS: $\Theta_{E1}$ (276.9 K) accounts for the presence of 20 high energy modes above the gap and $\Theta_{E2}$ (77.6 K) partially accounts for the accumulation of low-lying optical modes. The acoustic modes and part of the low-lying optical modes are described by the Debye and non-Debye contributions of model 2 (Table 1). Hence, the proposed Debye-Einstein model appears to be physically reasonable - also in comparison with the theoretical PDOS.

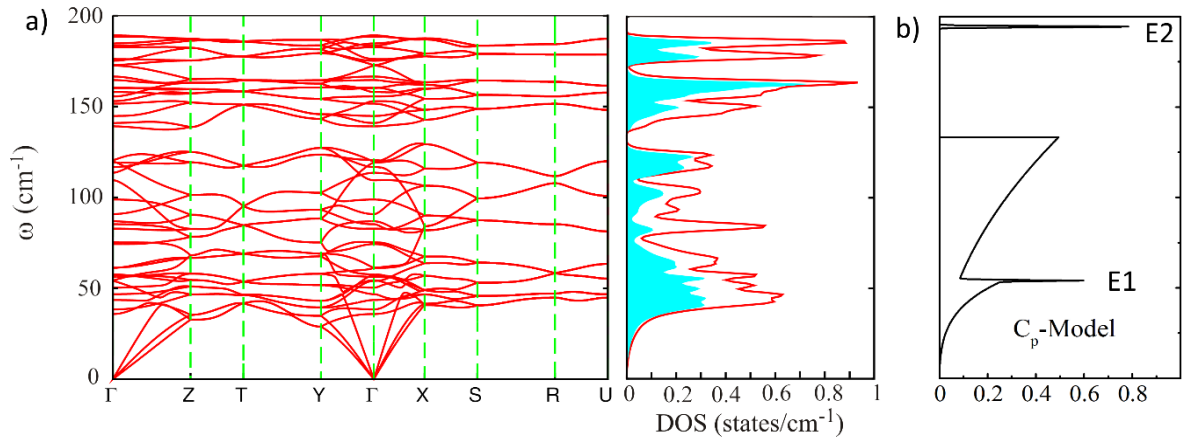

Figure 6. (a) Calculated phonon dispersion curves (left) and phonon density of states (PDOS) of ZnSb (right). The partial Sb atom contribution to the PDOS is indicated by the cyan area. (b) PDOS model based on the Debye and Einstein temperatures as extracted from the measured heat capacity. (c) PDOS constructed from fitting model 2 (cf. Table 1).

To probe the optic modes of ZnSb we performed Raman spectroscopy on single crystal and SPS consolidated specimens. The orthorhombic structure with eight formula units gives rise to 24 Raman active $A_g$, $B_{1g}$, $B_{2g}$, $B_{3g}$ modes, 15 IR active $B_{1u}$, $B_{2u}$ and $B_{3u}$ modes and 6 silent $A_u$ modes. An earlier Raman study by Smirnov *et al.* claimed the detection of 16 modes [42,43]. However most bands were very broad and some were very weak. It appears doubtful whether 16 modes were really detected. Furthermore their assignment is very difficult because of the similar wavenumbers. The general features of the Raman spectrum of ZnSb are an asymmetric band with high intensity at around 173 cm$^{-1}$ and several weak and broad bands below 80 cm$^{-1}$. This overall appearance was confirmed more recently by Trichês *et al.* [44].

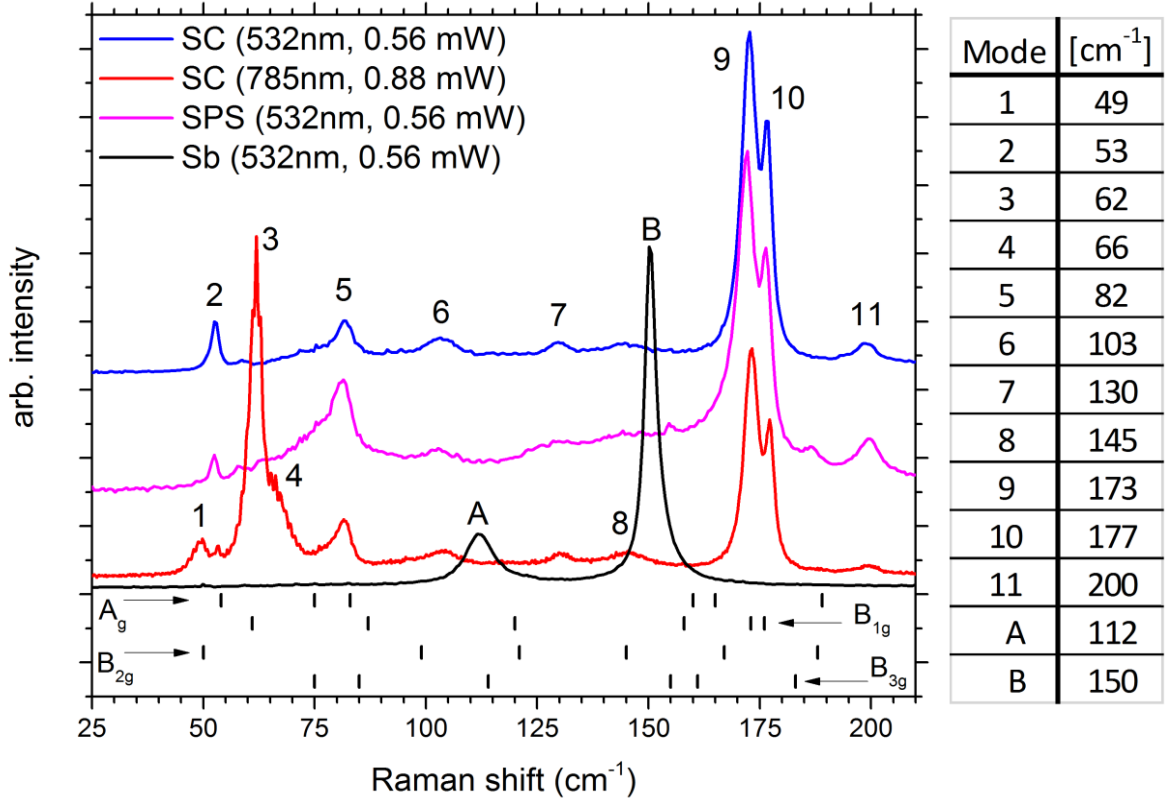

Figure 7: Raman spectrum of ZnSb and elemental antimony recorded with an excitation wavelength of 532 nm and 785 nm. The listed values on the right hand side are given in wavenumbers (cm$^{-1}$) for the labeled bands 1-11 for ZnSb and A-B for elemental antimony. The wavenumbers of the individual optical modes were obtained from the DFT calculation at the $\Gamma$–point and are shown as bars at the bottom of the spectra.

Our spectra are compiled in Fig. 7 and compared with the experimental spectrum of elemental Sb and the theoretical wavenumbers of the Raman active modes obtained from first principles calculated. Rhombohedral Sb displays two Raman active modes: the $A_g$ mode at 150 cm$^{-1}$ which is characterized by Sb atoms shifting along the $C_3$ axis, and the degenerate $E_g$ band at 110 cm$^{-1}$ which triggers the displacement of atoms perpendicular to the $C_3$ axis. We note that the stretching mode of $Sb_2$ moieties in ZnSb occurs at higher wavenumbers (ca. 165 cm$^{-1}$). Compared to the previous measurements [41–43] our spectra display a lower signal to noise ratio and a higher resolution. To record well-resolved spectra it is salient to perform excitation with a green laser ($\lambda$ = 532 nm). In that case the spectra are characterized by a prominent band is at 173 cm$^{-1}$. However in contrast with the earlier measurements [41–43] this band is now clearly split into two components (173 cm$^{-1}$ and 173 cm$^{-1}$). Further bands are consistently observed at ca. 200 cm$^{-1}$, 75 – 80 cm$^{-1}$, and 50 – 55 cm$^{-1}$. Spectra from single crystal specimens display additional bands at ~100, ~130, and ~145 cm$^{-1}$. We note that the location of bands may vary by 1 – 3 cm$^{-1}$ in different spectra due to the local heating of the sample by the laser beam. In spectra obtained from red laser excitation ($\lambda$ = 785 nm) the low energy modes become more pronounced. In addition to the occurrence of a band at 50 cm$^{-1}$ we observe an additional intense band at around 60 cm$^{-1}$ with a characteristic shoulder at 65 cm$^{-1}$. The latter band could not be observed when the green laser excitation ($\lambda$ = 532 nm) was used instead. We further stress that it is important to excite the sample with a low laser power (0.56 mW, corresponding to a power density of approx. $5.5 \times 10^{-5}$ mW/μm$^2$). At higher laser powers bands at around 150 and 110 cm$^{-1}$ appear which signal the formation of elemental Sb and thus the onset of a thermally-induced decomposition of ZnSb in the laser beam.

In summary, our Raman investigation revealed up to 10 distinct bands in the range from 50 to 200 cm$^{-1}$ and unequivocally revealed the presence of low energy optic modes at around 50 cm$^{-1}$. Calculated wavenumbers for Raman active modes appear to be underestimated by 5-10%. Smirnov et al. also characterized the IR modes of ZnSb from reflectivity spectra and magnetophonon resonance data. [42] These IR modes were found in the ranges 184 – 195 cm$^{-1}$, 155 – 166 cm$^{-1}$, 119 – 123 cm$^{-1}$, and 44 – 66 cm$^{-1}$.

With knowledge of the PDOS of ZnSb, the theoretical vibrational heat capacity at constant volume ($C_v$) can be calculated according to Equation (3), c.f. Figure 4 and Figure 5. Following the analysis by Jund *et al.* [17], a linear term $A_1 = A_1^{QH} + A_1^{anh}$ may be added to the harmonically calculated $C_v(T)$ values (c.f. Equation 11) to account for the (quasi-harmonic, QH) volume change and anharmonic contributions, respectively (denoted as *A* and *B* in Ref. [17], respectively). The value of $A_1^{QH}$ can be approximated as $(B_M V \alpha_v^2 / C_{v,\infty})$ – where $B_M$ is the bulk modulus, $V$ is the molar volume, and $\alpha_v$ is the volume expansion coefficient. $A_1^{QH}$ is then obtained as $49.9 \times 10^{-6}$ K$^{-1}$ using values of $B_M$ from the literature ($B_M \sim 50$ GPa [17,41,44]) and values of $V$ and $\alpha_v^2$ from this work. The correction term $A_1 = A_1^{QH}$ was then applied to calculate $C_p$ (cf. Fig. 4). Figure 4 clearly reveals that in the high temperature range the experimental $C_p(T)$ values are slightly larger than the theoretical $C_p$ values (note that the $A_1$ parameters obtained from the Debye-Einstein fitting are substantially higher, cf. Table 1). For the single crystal $C_p$ data we attribute this discrepancy to anharmonicity (the corresponding value for $A_1^{anh}$ would then be $43.7 \times 10^{-6}$ K$^{-1}$), whereas for the SPS sample defects, stress or strain may provide additional contributions.

Below $T = 150$ K, the calculated $C_p$ values are higher than the experimental $C_p$ values. This deviation becomes even more pronounced with decreasing temperature (see inset in Fig. 4). This discrepancy is primarily attributed to an underestimation of the calculated phonon frequencies (as indicated from the comparison of calculated Raman modes and measured bands). At these low temperatures, underestimated phonon frequencies will lead to an overestimation of the heat capacity. Due to the asymptotic nature of $C_v$ with respect to the Dulong-Petit limit, this discrepancy is only of minor importance at elevated temperatures. As a consequence, the Debye and Einstein temperatures obtained from a fit of the calculated $C_p(T)$ data to a Debye-Einstein model (model 2), are considerably lower compared to the values of the experimental $C_p$ data (cf. Table 1).

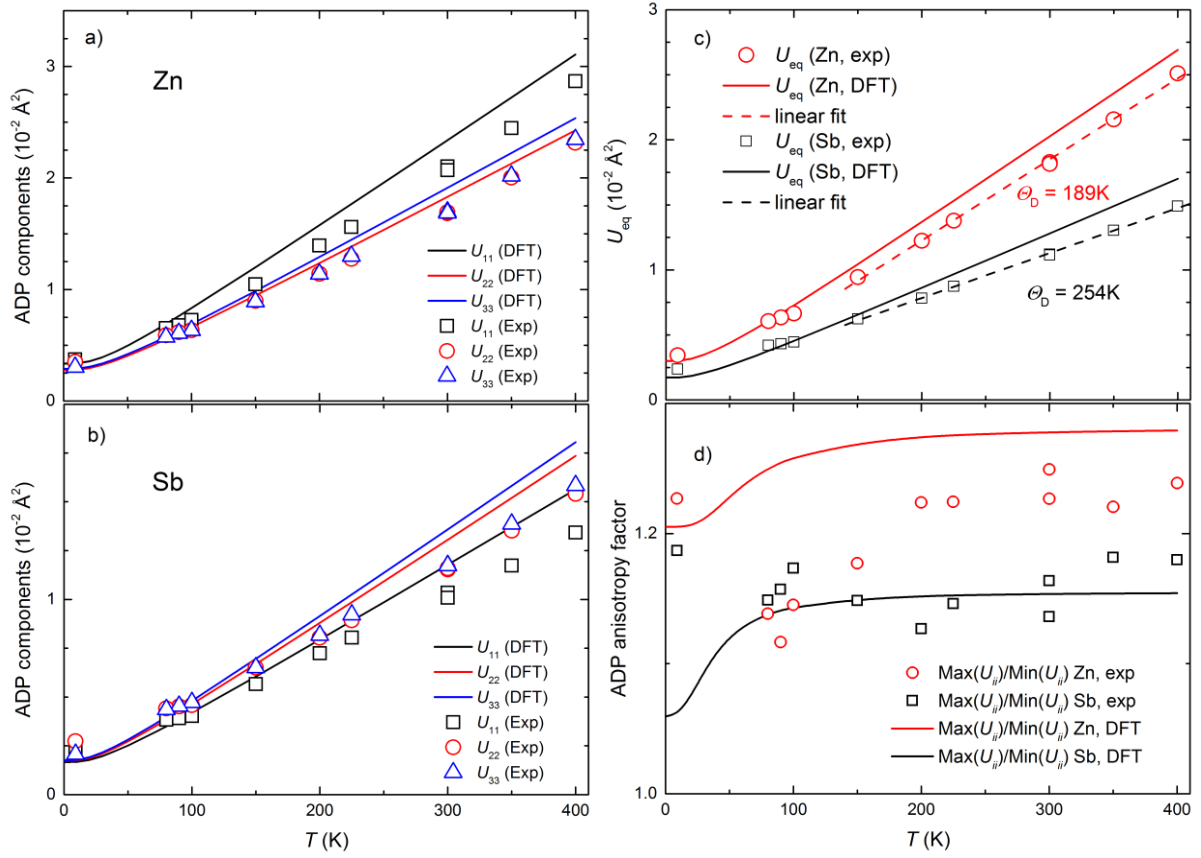

Figure 8. (a,b) Theoretical harmonic atomic displacement parameters (ADPs) obtained from DFT (solid lines) in comparison with experimental values, (c) comparison of $U_{eq}$ values, which are calculated as the mean of the diagonal elements of the harmonic ADP tensor, (d) comparison of the ADP anisotropy factor, which is defined as the ratio between maximum $U_{ii}$ and minimum $U_{ii}$ values. A ratio of 1 would correspond to an isotropic displacement.

In order to investigate in greater detail the nature of normal modes of vibrations in ZnSb we examined the temperature dependency of the ADPs of the Zn and Sb atoms obtained from refinements of an independent atom model against high resolution X-Ray data. In a perfect crystal harmonic ADPs represent the anisotropic temperature factors and describe the mean-square vibrational amplitude of an atom with respect to its equilibrium position in the crystal. Fig. 8ab compares the diagonal elements of the $U_{ij}$ tensor of the experimental ADPs and averaged $U_{iso}$ values with ones derived from DFT calculations (Eq. 5 – 8). The ratio of the calculated diagonal elements is qualitatively in agreement with experiment (i.e. for the zink atoms: $U_{11} > U_{22} \approx U_{33}$ and for antimony atoms: $U_{33} \approx U_{22} > U_{11}$). The trend in temperature dependency of the absolute values of the $U_{ij}$ tensor are also reproduced by the theoretical results. However, the calculated displacement parameters are slightly larger than the experimental ones at higher temperatures. ADP analysis reveals that the $U_{ii}$ values of the Zn atom are in general larger than the corresponding ones of the antimony atoms at the same temperature and the show a larger increase in magnitude with increasing temperature relative to the Sb atoms. This trend is consistent between the theoretical and experimental results and is therefore not significantly biased by a potential disorder of the Zn atoms which would affect the experimental data only (which actually displays smaller $U_{ii}$ values compared to the calculated ones by DFT). A severe deficiency of Zn atoms as another potential origin of the large $U_{ii}$ values of Zn can be also ruled out since the refinements against the high resolution diffraction data does not reveal any statistically significant deviation of the Zn atom site from full occupancy. Analysis of the ADPs obtained by experiment and the DFT calculation, however, cannot be used to furnish or preclude the presence of a deficient site occupancy of Zn due to defects. Indeed, reduction of the fractional site occupancy of Zn by 0.5% (which corresponds to an unrealistically high defect concentration of $1.03 \times 10^{20}$ cm$^{-3}$) causes a reduction of the corresponding $U_{iso}$(Zn) values by merely 0.7 %, - a value which is below statistical significance in the refinements of our X-ray data. Accordingly, the hypothesis that that electrical properties of ZnSb are largely influenced by the presence of Zn defects [6,45] can neither be confirmed nor ruled out on the basis of the X-ray data alone. Also the presence of a severe substitutional Zn/Sb disorder is not indicated by our refinements. The latter case would be clearly indicated by an erroneous decrease of the $U_{ii}$(Zn) in the refinements and the presence of large residual densities in the core region of the zinc atoms [46]. Hence, the temperature dependent trend in the $U_{ii}$ values is mainly attributed to the fact that the Zn atoms display a significantly smaller mass than the antimony atoms (most important isotopes: $^{64}$Zn and $^{121}$Sb). As a consequence, the theoretical $U_{eq}$ value of the zinc atom which is due to zero point motion is already significantly larger (about 0.001 Å$^2$) than the corresponding $U_{iso}$(Sb) value at 0K. This difference becomes even more pronounced at elevated temperature and reaches 0.01 Å$^2$ at 400 K.

It is common practice for the analysis of thermoelectric materials to use the slope of the individual $U_{eq}(T)$ sequences (c.f. Figure 8c) to derive Debye and Einstein temperatures [47,48]. For the Debye case and $T > \Theta_D$ the temperature dependence of $U_{eq}(T)$ corresponds to:

$$U_{eq}(T) = \frac{3h^2 T}{4\pi^2 m_{at} k_B \Theta_D^2} \qquad (12)$$

,where $m_{at}$ is the mass of the respective atoms in ZnSb. The derived Debye temperatures $\Theta_{D,exp,ADP}$ using experimental $U_{eq}(T)$ values above 150K for ZnSb are 189K and 254K for Zn and Sb atoms, respectively. In case of the DFT derived $U_{eq}(T)$ values the corresponding $\Theta_{D,theo,ADP}$ values are 183K and 230K for Zn and Sb atoms, respectively. We note that the results for the Sb atoms compare reasonably well with the overall Debye temperature $\Theta_D(0)$ from specific heat analysis (248K from experiment and 210K from DFT), while for Zn atoms the values are significantly lower. However, the difference between theory and experiment is more pronounced for Sb atoms.

In addition the $U_{ii}$ values of Zn atoms also increase in anisotropy which is shown by the ratio of minimum and maximum $U_{ii}$ values, cf. Figure 8d. The anisotropy for the DFT result is even higher, but less temperature dependent above 80K. For Sb atoms the anisotropy is rather constant and agrees well with the DFT prediction.

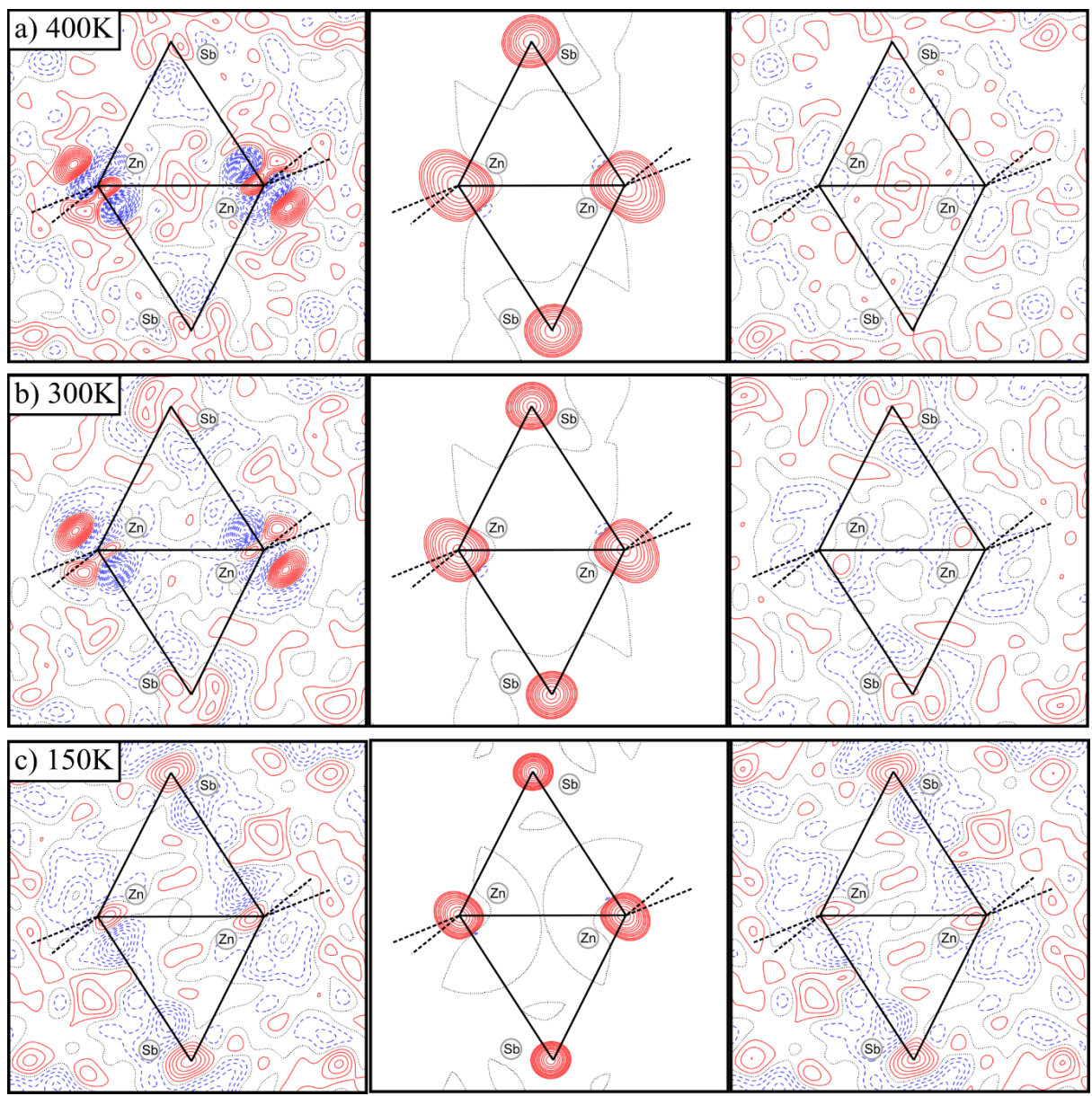

Figure 9. Difference electron density before (left) and after (right) anharmonic refinement for temperatures of 400 K, 300 K and 150 K. The positive (red) and negative (blue) contour values are shown in 0.2e/Å³ steps. The probability density distribution (middle) is given with contour values ($2, 4, 8 \times 10^n$, $n = -2 \ldots 3$).

When analyzing the residual density distributions of the harmonic refinements we realized significant residual density features of positive and negative sign in alternating order at the Zn positions, which is a typical indicator for anharmonic motion [49]. This is shown in Fig. 9 (left panel). Third order (and corresponding higher order) ADPs, commonly known as the Gram-Charlier expansion are used describe this anharmonic motion. In case of ZnSb refinements of Gram-Charlier expansion of orders higher than 3 did not improve the flatness of the residual density maps significantly and were not pursued. Accordingly, only the third order anharmonic parameters were refined to model the ADPs of the zinc and antimony atoms. Since both Zn and Sb occupy general crystallographic positions, 10 third order anharmonic parameters per atom need to be considered. Their refinements helped to improve the data fit dramatically as evidenced by the further decrease of the R-value from 2.08 to 1.58 % for the data collected at 400 K. As demonstrated in Fig. 9, also the resulting residual electron density maps became essentially flat as a consequence of these anharmonic refinements. Analyses of the resulting probability density distributions do not reveal any pronounced negative regions which is a prerequisite of a physically valid model [50].

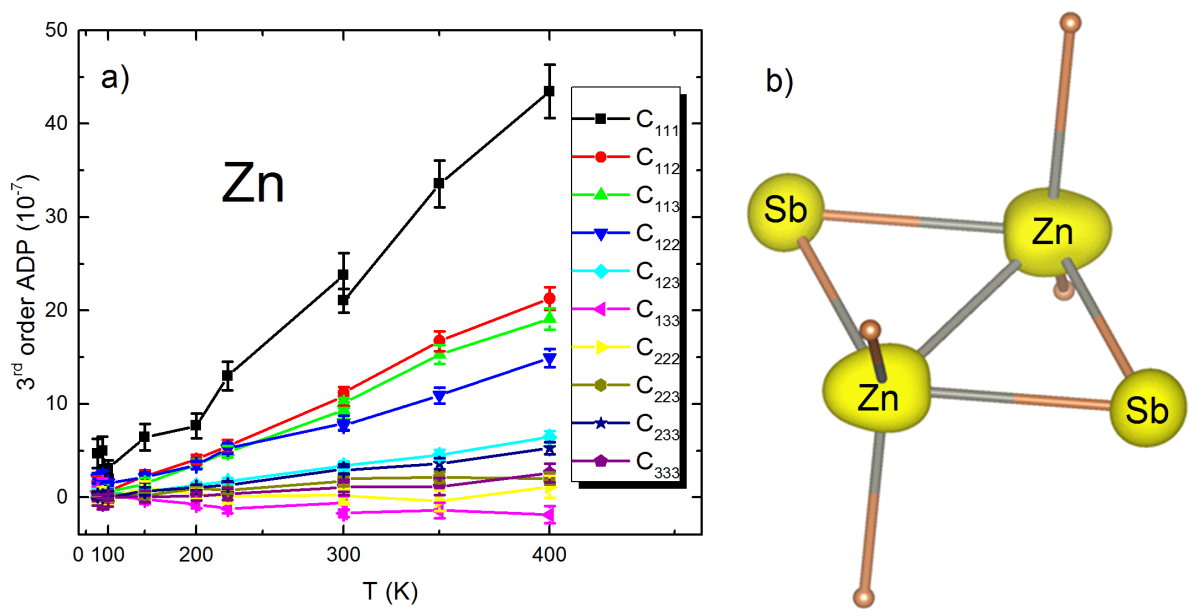

Figure 10. Third order Gram-Charlier ADP expansion coefficients for Zn (a) versus $T$ ($T$-axis scaled as $T^2$). Estimated standard uncertainties of the refined parameters are shown as error bars and b) isosurface (surface value 0.1) representation of the nuclear probability density of the $Zn_2Sb_2$ entity at 400 K.

Interestingly, only the Zn atoms are characterized by a significant anharmonicity of its vibrational amplitudes. The temperature-dependency of the third order Gram-Charlier expansion coefficients for Zn are shown in Fig. 10a. In case of a truly anharmonic vibrational behaviour the $C_{ijk}$ paramters should scale as $T^2$, which is the case for Zn. For Sb all anharmonic parameters vanish within their standard uncertainty (see Supporting Information). The nuclear probability density of a $Zn_2Sb_2$ unit at 400 K is depicted in Fig. 10b. We note that the refinement of anharmonic ADP has almost no influence on the values of the harmonic ADPs, in agreement with recent findings from theoretical models [51], and has only little influence on the interatomic distances (taken into account for in Fig. 2c). Most notably, the Zn–Zn distance increases by 0.5% and the r2 distance by 0.3% compared to the purely harmonic model at a temperature of 400K.

In conclusion, the refinement of anharmonic ADPs clearly validates the presence of anharmonicity in ZnSb. Moreover, the anharmonic behavior affects only the Zn atoms. Anharmonicity was already indicated from the analysis of the heat capacity and accounted for by introducing a $A_1^{anh}$ parameter.

## D. Origin of low thermal conductivity

The lattice thermal conductivity of ZnSb is remarkably low [6,52] and is characterized for temperatures above 200 K by values lower than 2 W/mK. This characteristic behavior of ZnSb compares well to bulk PbTe which is one of the most established and approved thermoelectric materials. At first sight rocksalt structured PbTe, which exclusively constitutes of heavy atoms, and ZnSb do not have much in common neither with regard to chemical composition nor with respect to structural chemistry. However, we argue that the physical mechanism which is responsible for the low lattice thermal conductivity of both materials should be strikingly similar, despite their distinct differences in composition and structure-property relationships.

Recent inelastic neutron scattering experiments on PbTe revealed a peculiar dynamic behavior: Through anharmonic coupling the ferroelectric transverse optic (TO) mode interacts with the heat carrying longitudinal acoustic phonons over a wide range of frequencies, thus resembling a rattling mode [53]. Other studies showed disorder for the Pb atoms, which displace increasingly with temperature from their ideal position in the rocksalt structure [54,55]. The dynamical peculiarity and structural disorder are certainly correlated and have been interpreted by considering PbTe as an incipient ferroelectric material [53]. The incipient ferroelectric state in turn is a consequence of Pb 6p-Te 5p bonding.

We propose, that the role of the ferroelectric TO mode in PbTe is adopted weakly dispersed low-energy optic modes in ZnSb. Their presence is a natural consequence of localized multicenter bonding established within the rhomboid $Zn_2Sb_2$ rings. The peculiar feature of the PDOS of ZnSb is the accumulation of 12 optic modes in a narrow window between 36 and 61 cm$^{-1}$. This assembly of 12 soft modes possesses a great flexibility, which allows for effective interaction with heat carrying acoustic phonons throughout large parts of the Brillouin zone (cf. Fig. 6). The presence of these soft modes is also reflected in the increasing anharmonicity of the mean-square vibrational amplitudes of the zinc atoms with rising temperature in the diffraction experiments. Accordingly, the presence of anharmonic vibrational behavior and the delocalized character of the electron-poor 4c4e bonds in the $Zn_2Sb_2$ moieties might represent two sides of the same coin and are finally responsible for the establishment of pronounced phonon-phonon interactions. Firstly, this scenario allows for additional (symmetry forbidden) interaction of optic and acoustic modes by anharmonic coupling. Secondly it promotes phonon-phonon scattering (Umklapp processes) at higher temperatures (typically above $\theta_D$).

In conclusion, there appears to be a great analogy in the individual mechanism causing the low lattice thermal conductivity for chemically PbTe and ZnSb – despite their distinct chemical composition and structures. Moreover, we believe that the peculiar phononic structure of ZnSb (displaying a multitude of localized low-energy modes, which can act as rattling modes toward heat carrying acoustic phonons) represents the sought-after underlying characteristic that accounts generally for an inherent low lattice thermal conductivity for zinc antimony compound and their derivatives. The presence of localized low energy optic modes is seen as a consequence of multicenter bonded structural entities which are common to all these compounds.

## IV. Conclusion

ZnSb possesses an inherently low lattice thermal conductivity, which is related in nature and magnitude to the one observed in the state-of-the-art thermoelectric material PbTe. From a combination of high resolution X-ray diffraction studies, Raman spectroscopy, heat capacity measurements and first principles calculations of the electronic and phononic structure we identify peculiar electronic and vibrational properties as control parameters of the thermoelectric properties of ZnSb. These properties manifest themselves in a multitude of localized low energy optic modes (which couple with the acoustic, heat carrying phonons) and the anharmonic vibrational behavior of the Zn atoms which are a natural consequence of the electron-poor character of the multicenter bonded structural $Zn_2Sb_2$ entities. Multicenter bonded structural entities are a common feature of thermoelectric electron poor semiconductors. We therefore argue that the vibrational behavior of ZnSb directly correlates with its bonding properties and that the established mechanisms for low thermal conductivity can be extended to other thermoelectric electron poor semiconductors like $Zn_4Sb_3$, $CdSb$, $Cd_4Sb_3$, $Cd_{13-x}In_yZn_{10}$, and $Zn_5Sb_4In_{2-\delta}$.

**Supporting Information Available.** Tables on the temperature dependent PXRD/SXRD measurements and refinements.

**Acknowledgements.** This work was supported by the US National Science Foundation (NSF-DMR-1007557), the Swedish Research Council (2010-4827 and 2013-4690) and the Deutsche Forschungsgemeinschaft (DFG, SCHE 478/12-1).

# Supporting Information for

# Thermal and Vibrational Properties of Thermoelectric ZnSb – Exploring the Origin of Low Thermal Conductivity

A. Fischer,[1] E.-W. Scheidt,[1] W. Scherer,[1] D. Benson,[2] Y. Wu,[3,†] D. Eklöf,[4] U. Häussermann[4]

Table of Contents

**S1: Crystallographic Tables for PXRD refinements**

| material | ZnSb | ZnSb | ZnSb | ZnSb | ZnSb | ZnSb |
|---|---|---|---|---|---|---|
| $T$ [K] | 300(2) | 323(2) | 373(2) | 423(2) | 473(2) | 523(2) |
| ProfileScale | 1.103(5) | 1.103(5) | 1.100(5) | 1.097(5) | 1.088(6) | 1.082(6) |
| ZeroShift | -0.55(9) | -1.29(9) | -2.51(9) | -4.26(9) | -6.60(9) | -8.56(9) |
| $a$ [Å] | 6.2055(2) | 6.2084(2) | 6.2150(2) | 6.2223(2) | 6.2301(2) | 6.2373(2) |
| $b$ [Å] | 7.7460(2) | 7.7479(2) | 7.7518(2) | 7.7563(2) | 7.7620(2) | 7.7664(2) |
| $c$ [Å] | 8.1026(2) | 8.1035(2) | 8.1042(2) | 8.1058(2) | 8.1080(2) | 8.1095(2) |
| $U_{iso}$(Zn) [Å$^2$] | 0.018* | 0.020* | 0.021* | 0.026* | 0.029* | 0.033* |
| $U_{iso}$(Sb) [Å$^2$] | 0.012(1) | 0.013(1) | 0.014(1) | 0.017(1) | 0.019(1) | 0.021(2) |
| $LX$-parameter | 4.0(2) | 4.3(2) | 4.4(2) | 4.3(2) | 3.9(2) | 3.7(2) |
| $LY$-parameter | 2.5(7) | 1.7(7) | 1.5(7) | 1.1(7) | 0.8(7) | 0.2(7) |
| $GW$-parameter | 5.0(3) | 4.9(3) | 4.8(3) | 5.0(3) | 5.5(3) | 6.1(3) |
| $R_F$ [%] | 4.36 | 4.43 | 4.81 | 4.62 | 5.32 | 5.56 |
| $R_p$ / $R_{wp}$ [%] | 5.04/6.52 | 4.95/6.51 | 5.07/6.62 | 5.09/6.54 | 5.33/6.83 | 5.35/6.90 |
| $N_{param}$ (profile) | 20# | 20# | 20# | 20# | 20# | 20# |
| $N_{param}$ (struct.) | 7 | 7 | 7 | 7 | 7 | 7 |
| $2\vartheta_{max}$ [°] | 55 | 55 | 55 | 55 | 55 | 55 |
| GoF (=$\chi$) | 1.83 | 1.82 | 1.85 | 1.83 | 1.91 | 1.93 |

| material | ZnSb/Sb | ZnSb/Sb | ZnSb/Sb | ZnSb/Sb |
|---|---|---|---|---|
| $T$ [K] | 573(2) | 623(2) | 673(2) | 723(2) |
| ProfileScale | 1.043(5) | 1.024(5) | 0.994(5) | 0.868(5) |
| ZeroShift | -10.86(9) | -12.82(9) | -15.7(1) | -18.4(1) |
| $a$ [Å] (ZnSb) | 6.2449(2) | 6.2526(2) | 6.2612(2) | 6.2695(5) |
| $b$ [Å] (ZnSb) | 7.7717(2) | 7.7757(2) | 7.7817(2) | 7.7879(6) |
| $c$ [Å] (ZnSb) | 8.1113(2) | 8.1123(2) | 8.1147(2) | 8.1186(7) |
| $a$ [Å] (Sb) | 4.321(1) | 4.3229(6) | 4.3245(3) | 4.3268(1) |
| $c$ [Å] (Sb) | 11.353(6) | 11.353(3) | 11.361(1) | 11.3669(4) |
| $U_{iso}$(Zn) [Å$^2$] | 0.030* | 0.029* | 0.027* | 0.027* |
| $U_{iso}$(Sb) [Å$^2$] | 0.019(1) | 0.018(1) | 0.017(1) | 0.017* |
| $LX$-parameter (ZnSb/Sb) | 3.3(1)/0* | 3.4(1)/0* | 3.2(1)/4.6(9) | 0.8(7) |
| $LY$-parameter (ZnSb/Sb) | 0* | 0* | 0* | 0* |
| $GW$-parameter (ZnSb/Sb) | 7.0(3)/18(6) | 6.5(3)/25(4) | 7.3(3)/10(3) | 14(2)/6.9(3) |
| phase fraction Sb [wt%] | 1(1) | 3(1) | 6.7(2) | 78.4(6) |
| $R_F$ [%] (ZnSb/Sb) | 5.63/ 4.48 | 5.56/4.60 | 6.57/3.80 | 21.16/2.40 |
| $R_p$ / $R_{wp}$ [%] | 5.09/ 6.60 | 4.89/6.55 | 4.68/6.46 | 5.15/7.27 |
| $N_{param}$ (profile) | 24# | 24# | 25# | 25# |
| $N_{param}$ (struct.) | 7 | 7 | 7 | 6 |
| $2\vartheta_{max}$ [°] | 55 | 55 | 55 | 55 |
| GoF (=$\chi$) | 1.84 | 1.82 | 1.80 | 2.09 |

\* not refined/constrained
# 12 parameters are used for the 12$^{th}$ order background Legendre-polynomial

## S2: Rietveld Refinement Details

For the refinement of the ADP (atomic displacement parameters) of ZnSb, the ratio $U_{iso}(Zn)/U_{iso}(Sb)$ was constrained for convergence reasons to the ratio of 1.56 in accordance with DFT calculations. In order to match their absolute values with single crystal X-ray data, an absorption coefficient of $\mu \cdot r = 0.034$ was employed. For the refinement of hexagonal, elemental Antimony (SG: $R\text{-}3m$), its $U_{iso}$ value was set to the value of antimony in ZnSb for convergence reasons. At the last temperature point ($T$ = 723K) the ADPs of ZnSb were kept fixed to their values at $T$ = 673K, while $U_{iso}$ of elemental antimony was freely refined.

## S3: Crystallographic Tables for SXRD refinements

For the analysis of the temperature dependent SXRD refinements we used a total of 4 different crystals in order to exclude any crystal dependent bias.

| Temperature [K] | Crystal |
|---|---|
| 9 | crystal 1 |
| 80 | crystal 2 |
| 90 | crystal 2 |
| 100 | crystal 3 |
| 150 | crystal 3 |
| 200 | crystal 3 |
| 225 | crystal 3 |
| 300 | crystal 1, crystal 4 |
| 350 | crystal 2 |
| 400 | crystal 2 |

|  | ZnSb 9K (crystal 1) | |
| --- | --- | --- |
| model | harmonic | anharmonic |
| formula | ZnSb | |
| crystal system | orthorhombic | |
| space group | *Pbca*, # 61 ITC | |
| *a* [Å] | 6.1616(4) | |
| *b* [Å] | 7.7133(5) | |
| *c* [Å] | 8.0828(5) | |
| *V* [Å³] | 384.15(4) | |
| $\rho$ [g/cm³] | 6.4712 | |
| *Z* | 8 | |
| *F*000 | 648 | |
| crystal size [µm³] | 84x125x135 | |
| $\mu$(Mo K$\alpha$) [mm$^{-1}$] | 25.99 | |
| *T* [K] | 9 (2) | |
| radiation | Mo K$\alpha$ ($\lambda$ = 0.71069 Å) | |
| $\theta$ range [°deg] | 4.93< $\theta$ <45.01 | |
| $(\sin\theta/\lambda)_{max}$ [Å$^{-1}$] | 0.95 | |
| data collected (*h*, *k*, *l*) | -10<*h*<10,-14<*k*<15,-16<*l*<16 | |
| no. of reflns measured | 14640 | |
| no. of unique reflns; | 1527 | |
| no. of observed reflns, *I*> 3$\sigma$(*I*) | 1348 | |
| no. refined parameters | 20 | |
| extinction type | Type 2 isotropic | |
| $\rho_{iso}$ (extinction) | 0.12(6) | |
| $R_{int}$ | 3.87 | |
| $R_1$ *I* > 3σ(*I*) [%] | 2.36 | |
| w$R_1$ *I* > 3σ(*I*) [%] | 2.57 | |
| GOF on F | 1.59 | |
| largest peak/hole [e/Å³] | 2.23/-2.85 | |
| absorption correction | numerical, face-indexed | |
| transmission min, max | 0.0968, 0.2288 | |

|  | ZnSb 80K (crystal 2) | |
| --- | --- | --- |
| model | harmonic | anharmonic |
| formula | ZnSb | |
| crystal system | orthorhombic | |
| space group | *Pbca*, # 61 ITC | |
| *a* [Å] | 6.1814(5) | |
| *b* [Å] | 7.7386(6) | |
| *c* [Å] | 8.1038(6) | |
| *V* [Å³] | 387.65(5) | |
| $\rho$ [g/cm³] | 6.4128 | |
| Z | 8 | |
| *F*000 | 648 | |
| crystal size [μm³] | 69x81x94 | |
| $\mu$(Ag Kα) [mm$^{-1}$] | 13.559 | |
| *T* [K] | 80(2) | |
| radiation | Ag Kα ($\lambda$ = 0.56087 Å) | |
| $\theta$ range [°deg] | 5.2<2$\theta$<72.2 | |
| (sin$\theta$/$\lambda$)$_{max}$ [Å$^{-1}$] | 1.06 | |
| data collected (*h, k, l*) | -12<*h*<12,-15<*k*<16,-16<*l*<17 | |
| no. of reflns measured | 17286 | |
| no. of unique reflns; | 1858 | |
| no. of observed reflns, *I*> 3σ(*I*) | 1540 | |
| no. refined parameters | 20 | 40 |
| extinction type | Type 2 isotropic | |
| $\rho_{iso}$ (extinction) | 0.18(1) | 0.18(1) |
| *R*$_{int}$ | 4.11 | |
| *R*$_1$ *I* > 3σ(*I*) [%] | 1.49 | 1.48 |
| w*R*$_1$ *I* > 3σ(*I*) [%] | 1.83 | 1.81 |
| GOF on F | 1.67 | 1.66 |
| largest peak/hole [e/Å³] | 1.46/-1.45 | 1.43/-1.49 |
| absorption correction | numerical, face-indexed | |
| transmission min, max | 0.4433,0.5762 | |

|  | ZnSb 90K (crystal 2) | |
| --- | --- | --- |
| model | harmonic | anharmonic |
| formula | ZnSb | |
| crystal system | orthorhombic | |
| space group | *Pbca*, # 61 ITC | |
| $a$ [Å] | 6.1818(4) | |
| $b$ [Å] | 7.7368(5) | |
| $c$ [Å] | 8.1021(6) | |
| $V$ [Å³] | 387.50(5) | |
| $\rho$ [g/cm³] | 6.4152 | |
| $Z$ | 8 | |
| $F000$ | 648 | |
| crystal size [µm³] | 62x78x109 | |
| $\mu$(Ag K$\alpha$) [mm$^{-1}$] | 13.564 | |
| $T$ [K] | 90(2) | |
| radiation | Ag K$\alpha$ ($\lambda$ = 0.56087 Å) | |
| $\theta$ range [°deg] | 5.2<2$\theta$<72.2 | |
| $(\sin\theta/\lambda)_{max}$ [Å$^{-1}$] | 1.06 | |
| data collected ($h$, $k$, $l$) | -12<$h$<12,-16<$k$<15,-16<$l$<16 | |
| no. of reflns measured | 16660 | |
| no. of unique reflns; | 1829 | |
| no. of observed reflns, $I$> 3$\sigma$($I$) | 1539 | |
| no. refined parameters | 20 | 40 |
| extinction type | Type 2 isotropic | |
| $\rho_{iso}$ (extinction) | 0.18(1) | 0.19(1) |
| $R_{int}$ | 3.94 | |
| $R_1$ $I$ > 3$\sigma$($I$) [%] | 1.50 | 1.50 |
| w$R_1$ $I$ > 3$\sigma$($I$) [%] | 1.79 | 1.76 |
| GOF on F | 1.67 | 1.66 |
| largest peak/hole [e/Å³] | 1.81/-1.52 | 1.74/-1.46 |
| absorption correction | numerical, face-indexed | |
| transmission min, max | 0.4470,0.5711 | |

|  | ZnSb 100K (crystal 3) | |
| --- | --- | --- |
| model | harmonic | anharmonic |
| formula | ZnSb | |
| crystal system | orthorhombic | |
| space group | *Pbca*, # 61 ITC | |
| $a$ [Å] | 6.1838(3) | |
| $b$ [Å] | 7.7373(4) | |
| $c$ [Å] | 8.1080(4) | |
| $V$ [Å³] | 387.93(3) | |
| $\rho$ [g/cm³] | 6.408 | |
| $Z$ | 8 | |
| $F000$ | 648 | |
| crystal size [µm³] | 62x71x79 | |
| $\mu$(Ag K$\alpha$) [mm$^{-1}$] | 13.549 | |
| $T$ [K] | 100(2) | |
| radiation | Ag K$\alpha$ ($\lambda$ = 0.56087 Å) | |
| $\theta$ range [°deg] | 5.2<2$\theta$<72.2 | |
| $(\sin\theta/\lambda)_{max}$ [Å$^{-1}$] | 1.31 | |
| data collected ($h$, $k$, $l$) | -15<$h$<16,-20<$k$<20,-21<$l$<21 | |
| no. of reflns measured | 41895 | |
| no. of unique reflns; | 3687 | |
| no. of observed reflns, $I> 3\sigma(I)$ | 1603 | |
| no. refined parameters | 20 | 40 |
| extinction type | Type 2 isotropic | |
| $\rho_{iso}$ (extinction) | 0.31(6) | 0.31(6) |
| $R_{int}$ | 2.62 | |
| $R_1$ $I > 3\sigma(I)$ [%] | 1.01 | 1.00 |
| w$R_1$ $I > 3\sigma(I)$ [%] | 1.22 | 1.20 |
| GOF on F | 2.40 | 2.37 |
| largest peak/hole [e/Å³] | 1.33/-1.33 | 1.33/ -1.31 |
| absorption correction | numerical, face-indexed | |
| transmission min, max | 0.4498,0.6374 | |

|  | ZnSb 150K (crystal 3) | |
| --- | --- | --- |
| model | harmonic | anharmonic |
| formula | ZnSb | |
| crystal system | orthorhombic | |
| space group | *Pbca*, # 61 ITC | |
| $a$ [Å] | 6.1937(4) | |
| $b$ [Å] | 7.7443(5) | |
| $c$ [Å] | 8.1070(5) | |
| $V$ [Å³] | 388.86(4) | |
| $\rho$ [g/cm³] | 6.3928 | |
| $Z$ | 8 | |
| $F000$ | 648 | |
| crystal size [μm³] | 62x63x88 | |
| $\mu$(Ag Kα) [mm$^{-1}$] | 13.516 | |
| $T$ [K] | 150(2) | |
| radiation | Ag Kα ($\lambda$ = 0.56087 Å) | |
| $\theta$ range [°deg] | 3.87< $\theta$ < 32.23 | |
| $(\sin\theta/\lambda)_{max}$ [Å$^{-1}$] | 0.95 | |
| data collected ($h, k, l$) | -11<$h$<11,-14<$k$<14,-15<$l$<13 | |
| no. of reflns measured | 10050 | |
| no. of unique reflns; | 1344 | |
| no. of observed reflns, $I$> 3σ($I$) | 1294 | |
| no. refined parameters | 20 | 40 |
| extinction type | Type 2 isotropic | |
| $\rho_{iso}$ (extinction) | 0.41(7) | 0.41(7) |
| $R_{int}$ | 2.59 | |
| $R_1$ $I$ > 3σ($I$) [%] | 1.25 | 1.24 |
| w$R_1$ $I$ > 3σ($I$) [%] | 1.42 | 1.38 |
| GOF on F | 1.85 | 1.82 |
| largest peak/hole [e/Å³] | 1.14/-1.40 | 1.12/-1.21 |
| absorption correction | numerical, face-indexed | |
| transmission min, max | 0.4543,0.6251 | |

|  | ZnSb 200K (crystal 3) | |
| --- | --- | --- |
| model | harmonic | anharmonic |
| formula | ZnSb | |
| crystal system | orthorhombic | |
| space group | *Pbca*, # 61 ITC | |
| $a$ [Å] | 6.1995(4) | |
| $b$ [Å] | 7.7469(5) | |
| $c$ [Å] | 8.1117(5) | |
| $V$ [Å³] | 389.58(4) | |
| $\rho$ [g/cm³] | 6.381 | |
| $Z$ | 8 | |
| $F000$ | 648 | |
| crystal size [μm³] | 64x69x84 | |
| $\mu$(Ag Kα) [mm$^{-1}$] | 13.491 | |
| $T$ [K] | 200(2) | |
| radiation | Ag Kα ($\lambda$ = 0.56087 Å) | |
| $\theta$ range [°deg] | 3.87< $\theta$ <34.35 | |
| $(\sin\theta/\lambda)_{max}$ [Å$^{-1}$] | 1.00 | |
| data collected ($h, k, l$) | -11<$h$<11,-15<$k$<15,-14<$l$<16 | |
| no. of reflns measured | 10353 | |
| no. of unique reflns; | 1594 | |
| no. of observed reflns, $I> 3\sigma(I)$ | 1522 | |
| no. refined parameters | 20 | 40 |
| extinction type | Type 2 isotropic | |
| $\rho_{iso}$ (extinction) | 0.44(8) | 0.44(8) |
| $R_{int}$ | 2.51 | |
| $R_1$ $I > 3\sigma(I)$ [%] | 1.34 | 1.24 |
| w$R_1$ $I > 3\sigma(I)$ [%] | 1.48 | 1.38 |
| GOF on F | 1.88 | 1.76 |
| largest peak/hole [e/Å³] | 1.11/-1.66 | 1.10/-1.36 |
| absorption correction | numerical, face-indexed | |
| transmission min, max | 0.4155,0.6061 | |

| | ZnSb 225K (crystal 3) | |
|---|---|---|
| model | harmonic | anharmonic |
| formula | ZnSb | |
| crystal system | orthorhombic | |
| space group | *Pbca*, # 61 ITC | |
| $a$ [Å] | 6.2028(3) | |
| $b$ [Å] | 7.7490(4) | |
| $c$ [Å] | 8.1114(4) | |
| $V$ [Å³] | 389.88(3) | |
| $\rho$ [g/cm³] | 6.3761 | |
| $Z$ | 8 | |
| $F000$ | 648 | |
| crystal size [μm³] | 58x62x70 | |
| $\mu$(Ag Kα) [mm$^{-1}$] | 13.481 | |
| $T$ [K] | 225(2) | |
| radiation | Ag Kα ($\lambda$ = 0.56087 Å) | |
| $\theta$ range [°deg] | 3.87< $\theta$ < 32.19 | |
| $(\sin\theta/\lambda)_{max}$ [Å$^{-1}$] | 0.95 | |
| data collected ($h, k, l$) | -11<$h$<11,-14<$k$<14,-13<$l$<15 | |
| no. of reflns measured | 10034 | |
| no. of unique reflns; | 1346 | |
| no. of observed reflns, $I > 3\sigma(I)$ | 1285 | |
| no. refined parameters | 20 | 40 |
| extinction type | Type 2 isotropic | |
| $\rho_{iso}$ (extinction) | 0.47(9) | 0.47(9) |
| $R_{int}$ | 2.64 | |
| $R_1$ $I > 3\sigma(I)$ [%] | 1.40 | 1.27 |
| w$R_1$ $I > 3\sigma(I)$ [%] | 1.57 | 1.40 |
| GOF on F | 2.05 | 1.85 |
| largest peak/hole [e/Å³] | 1.25/-1.81 | 1.19/-1.31 |
| absorption correction | numerical, face-indexed | |
| transmission min, max | 0.4770, 0.6500 | |

|  | ZnSb 300K (crystal 4) | |
| --- | --- | --- |
| model | harmonic | anharmonic |
| formula | ZnSb | |
| crystal system | orthorhombic | |
| space group | *Pbca*, # 61 ITC | |
| *a* [Å] | 6.2084(4) | |
| *b* [Å] | 7.7521(5) | |
| *c* [Å] | 8.1101(5) | |
| *V* [Å³] | 390.32(4) | |
| $\rho$ [g/cm³] | 6.3688 | |
| *Z* | 8 | |
| *F*000 | 648 | |
| crystal size [μm³] | 67x78x130 | |
| $\mu$(Ag Kα) [mm$^{-1}$] | 13.466 | |
| *T* [K] | 300 (2) | |
| radiation | Ag Kα ($\lambda$ = 0.56087 Å) | |
| $\theta$ range [°deg] | 3.87< $\theta$ <40.86 | |
| $(\sin\theta/\lambda)_{max}$ [Å$^{-1}$] | 1.16 | |
| data collected (*h*, *k*, *l*) | -14<*h*<14,-18<*k*<18,-18<*l*<18 | |
| no. of reflns measured | 45248 | |
| no. of unique reflns; | 2592 | |
| no. of observed reflns, *I*> 3$\sigma$(*I*) | 1549 | |
| no. refined parameters | 20 | 40 |
| extinction type | Type 2 isotropic | |
| $\rho_{iso}$ (extinction) | 0.34(12) | 0.34(9) |
| $R_{int}$ | 4.32 | |
| $R_1$ *I* > 3σ(*I*) [%] | 1.59 | 1.25 |
| w$R_1$ *I* > 3σ(*I*) [%] | 1.94 | 1.45 |
| GOF on F | 2.30 | 1.73 |
| largest peak/hole [e/Å³] | 2.37/-2.35 | 0.89/-0.85 |
| absorption correction | numerical, face-indexed | |
| transmission min, max | 0.3324,0.5946 | |

|  | ZnSb 300K (crystal 1) | |
| --- | --- | --- |
| model | harmonic | anharmonic |
| formula | ZnSb | |
| crystal system | orthorhombic | |
| space group | *Pbca*, # 61 ITC | |
| $a$ [Å] | 6.2041(2) | |
| $b$ [Å] | 7.7423(2) | |
| $c$ [Å] | 8.1022(3) | |
| $V$ [Å³] | 389.18(2) | |
| $\rho$ [g/cm³] | 6.3875 | |
| $Z$ | 8 | |
| $F000$ | 648 | |
| crystal size [μm³] | 84x125x135 | |
| $\mu$(Ag Kα) [mm$^{-1}$] | 13.505 | |
| $T$ [K] | 300 (2) | |
| radiation | Ag Kα ($\lambda$ = 0.56087 Å) | |
| $\theta$ range [°deg] | 3.87< $\theta$ <32.55 | |
| $(\sin\theta/\lambda)_{max}$ [Å$^{-1}$] | 0.95 | |
| data collected ($h, k, l$) | -11<$h$<11,-14<$k$<14,-15<$l$<15 | |
| no. of reflns measured | 15625 | |
| no. of unique reflns; | 1387 | |
| no. of observed reflns, $I> 3\sigma(I)$ | 1300 | |
| no. refined parameters | 20 | 40 |
| extinction type | Type 2 isotropic | |
| $\rho_{iso}$ (extinction) | 0.46(16) | 0.47(14) |
| $R_{int}$ | 3.87 | |
| $R_1$ $I > 3\sigma(I)$ [%] | 1.72 | 1.48 |
| w$R_1$ $I > 3\sigma(I)$ [%] | 1.83 | 1.57 |
| GOF on F | 2.03 | 1.76 |
| largest peak/hole [e/Å³] | 1.49/-1.70 | 1.19/-1.22 |
| absorption correction | numerical, face-indexed | |
| transmission min, max | 0.2865, 0.4765 | |

|  | ZnSb 350K (crystal 2) | |
| --- | --- | --- |
| model | harmonic | anharmonic |
| formula | ZnSb | |
| crystal system | orthorhombic | |
| space group | *Pbca*, # 61 ITC | |
| $a$ [Å] | 6.2133(6) | |
| $b$ [Å] | 7.7516(7) | |
| $c$ [Å] | 8.1065(7) | |
| $V$ [Å³] | 390.43(6) | |
| $\rho$ [g/cm³] | 6.367 | |
| $Z$ | 8 | |
| $F000$ | 648 | |
| crystal size [µm³] | 67x81x89 | |
| $\mu$(Ag Kα) [mm$^{-1}$] | 13.462 | |
| $T$ [K] | 350 (2) | |
| radiation | Ag Kα ($\lambda$ = 0.56087 Å) | |
| $\theta$ range [°deg] | 3.86< $\theta$ <36.3 | |
| $(\sin\theta/\lambda)_{max}$ [Å$^{-1}$] | 1.05 | |
| data collected ($h, k, l$) | -12<$h$<12,-16<$k$<16,-16<$l$<17 | |
| no. of reflns measured | 16957 | |
| no. of unique reflns; | 1857 | |
| no. of observed reflns, $I> 3\sigma(I)$ | 1450 | |
| no. refined parameters | 20 | 40 |
| extinction type | Type 2 isotropic | |
| $\rho_{iso}$ (extinction) |  | 0.19(1) |
| $R_{int}$ | 4.01 | |
| $R_1$ $I > 3\sigma(I)$ [%] | 2.05 | 1.65 |
| w$R_1$ $I > 3\sigma(I)$ [%] | 2.12 | 1.70 |
| GOF on F | 1.84 | 1.48 |
| largest peak/hole [e/Å³] | 2.47/-2.82 | 0.89/-1.34 |
| absorption correction | numerical, face-indexed | |
| transmission min, max | 0.4329, 0.5562 | |

|  | ZnSb 400K (crystal 2) | |
| --- | --- | --- |
| model | harmonic | anharmonic |
| formula | ZnSb | |
| crystal system | orthorhombic | |
| space group | *Pbca*, # 61 ITC | |
| $a$ [Å] | 6.2213(7) | |
| $b$ [Å] | 7.7568(9) | |
| $c$ [Å] | 8.1094(10) | |
| $V$ [Å³] | 391.34(8) | |
| $\rho$ [g/cm³] | 6.3523 | |
| $Z$ | 8 | |
| $F000$ | 648 | |
| crystal size [μm³] | 61x87x95 | |
| $\mu$(Ag Kα) [mm⁻¹] | 13.431 | |
| $T$ [K] | 400 (2) | |
| radiation | Ag Kα ($\lambda$ = 0.56087 Å) | |
| $\theta$ range [°deg] | 3.86< $\theta$<36.3 | |
| $(\sin\theta/\lambda)_{max}$ [Å⁻¹] | | |
| data collected ($h, k, l$) | -13<$h$<12,-16<$k$<16,-16<$l$<17 | |
| no. of reflns measured | 17434 | |
| no. of unique reflns; | 1888 | |
| no. of observed reflns, $I> 3\sigma(I)$ | 1404 | |
| no. refined parameters | 20 | 40 |
| extinction type | Type 2 isotropic | |
| $\rho_{iso}$ (extinction) | | 0.21(1) |
| $R_{int}$ | 4.11 | |
| $R_1$ $I > 3\sigma(I)$ [%] | 2.09 | 1.59 |
| w$R_1$ $I > 3\sigma(I)$ [%] | 2.25 | 1.71 |
| GOF on F | 1.92 | 1.47 |
| largest peak/hole [e/Å³] | 2.44/-2.92 | 0.77/-0.90 |
| absorption correction | numerical, face-indexed | |
| transmission min, max | 0.4124, 0.5583 | |

**S4: Tables of crystal coordinates, harmonic and anharmonic order ADPs for SXRD data**

| | **ZnSb 9K** | **ZnSb 80K** | **ZnSb 90K** | **ZnSb 100K** | **ZnSb 150K** | **ZnSb 200K** |
|---|---|---|---|---|---|---|
| atom | Sb | Sb | Sb | Sb | Sb | Sb |
| $x$ | 0.14310(2) | 0.14307(3) | 0.14307(3) | 0.14304(2) | 0.14306(3) | 0.14309(2) |
| $y$ | 0.082104(17) | 0.08211(3) | 0.08208(2) | 0.082123(17) | 0.08212(2) | 0.082129(17) |
| $z$ | 0.108339(16) | 0.10836(3) | 0.10834(3) | 0.108326(17) | 0.10831(2) | 0.108290(18) |
| $U_{eq}$ | 0.00237(4) | 0.00406(2) | 0.00424(2) | 0.004387(14) | 0.00617(2) | 0.007774(18) |
| $U_{11}$ | 0.00230(7) | 0.00370(4) | 0.00386(4) | 0.00397(2) | 0.00564(4) | 0.00721(3) |
| $U_{22}$ | 0.00273(6) | 0.00425(4) | 0.00443(4) | 0.00452(2) | 0.00645(4) | 0.00800(3) |
| $U_{33}$ | 0.00207(5) | 0.00422(4) | 0.00444(4) | 0.00466(2) | 0.00644(4) | 0.00810(3) |
| $U_{12}$ | -0.00002(4) | 0.00006(2) | 0.00007(2) | 0.000101(13) | 0.000164(19) | 0.000237(15) |
| $U_{13}$ | -0.00004(4) | 0.00002(2) | 0.00005(2) | 0.000065(12) | 0.000138(19) | 0.000189(16) |
| $U_{23}$ | -0.00002(3) | 0.00006(2) | 0.00006(2) | 0.000066(12) | 0.000085(19) | 0.000093(15) |
| $C_{111}$ | - | 0.00004(7) | 0.00001(7) | -0.00001(4) | 0.00005(6) | 0.00004(5) |
| $C_{112}$ | - | 0.00003(3) | 0.00002(3) | 0.000020(18) | 0.00003(3) | 0.00001(2) |
| $C_{113}$ | - | 0.00004(3) | 0.00005(3) | 0.000011(16) | 0.00002(3) | 0.00001(2) |
| $C_{122}$ | - | -0.00001(2) | 0.00003(2) | -0.000007(14) | 0.00002(2) | 0.000017(17) |
| $C_{123}$ | - | -0.000005(16) | 0.000002(16) | -0.000003(9) | -0.000008(14) | 0.000002(11) |
| $C_{133}$ | - | 0.00000(2) | -0.00001(2) | -0.000014(12) | 0.00000(2) | 0.000021(16) |
| $C_{222}$ | - | 0.00006(4) | 0.00004(4) | 0.00009(2) | 0.00008(4) | 0.00008(3) |
| $C_{223}$ | - | 0.000019(19) | 0.000009(18) | 0.000006(11) | 0.000010(18) | 0.000002(13) |
| $C_{233}$ | - | 0.000010(18) | 0.000008(18) | 0.000014(10) | -0.000013(19) | 0.000011(14) |
| $C_{333}$ | - | 0.00000(3) | 0.00000(3) | 0.000005(19) | -0.00003(3) | -0.00002(2) |
| atom | Zn | Zn | Zn | Zn | Zn | Zn |
| $x$ | 0.54192(5) | 0.54251(7) | 0.54258(6) | 0.54248(4) | 0.54290(6) | 0.54332(5) |
| $y$ | 0.60936(3) | 0.60959(5) | 0.60962(5) | 0.60960(3) | 0.60995(4) | 0.61023(3) |
| $z$ | 0.62944(3) | 0.62957(4) | 0.62955(4) | 0.62962(3) | 0.62975(4) | 0.62997(3) |
| $U_{eq}$ | 0.00343(6) | 0.00593(4) | 0.00627(4) | 0.00658(2) | 0.00942(4) | 0.01221(3) |
| $U_{11}$ | 0.00373(12) | 0.00642(7) | 0.00675(7) | 0.00718(4) | 0.01047(6) | 0.01385(6) |
| $U_{22}$ | 0.00351(9) | 0.00575(7) | 0.00607(6) | 0.00624(4) | 0.00892(6) | 0.01139(5) |
| $U_{33}$ | 0.00304(9) | 0.00561(7) | 0.00599(7) | 0.00632(4) | 0.00887(7) | 0.01139(5) |
| $U_{12}$ | 0.00034(8) | 0.00075(5) | 0.00078(5) | 0.00090(3) | 0.00142(4) | 0.00201(4) |
| $U_{13}$ | 0.00008(8) | 0.00042(5) | 0.00050(5) | 0.00058(3) | 0.00089(4) | 0.00130(4) |
| $U_{23}$ | 0.00007(6) | 0.00044(5) | 0.00054(5) | 0.00061(3) | 0.00102(4) | 0.00143(4) |
| $C_{111}$ | - | 0.00047(16) | 0.00050(15) | 0.00031(9) | 0.00064(14) | 0.00077(13) |
| $C_{112}$ | - | 0.00016(7) | 0.00012(6) | 0.00007(4) | 0.00023(6) | 0.00041(5) |
| $C_{113}$ | - | 0.00005(6) | 0.00009(6) | 0.00006(4) | 0.00015(6) | 0.00035(5) |
| $C_{122}$ | - | 0.00021(5) | 0.00024(5) | 0.00016(3) | 0.00022(5) | 0.00035(4) |
| $C_{123}$ | - | 0.00006(3) | 0.00004(3) | 0.000040(19) | 0.00006(3) | 0.00013(3) |
| $C_{133}$ | - | 0.00007(5) | 0.00016(5) | 0.00002(3) | -0.00002(5) | -0.00007(4) |
| $C_{222}$ | - | 0.00000(7) | 0.00004(7) | 0.00000(4) | 0.00010(7) | 0.00004(6) |
| $C_{223}$ | - | 0.00004(4) | 0.00000(4) | 0.00004(2) | 0.00002(4) | 0.00009(3) |
| $C_{233}$ | - | 0.00002(4) | 0.00001(4) | 0.00000(2) | 0.00007(4) | 0.00011(3) |
| $C_{333}$ | - | -0.00002(6) | -0.00007(6) | 0.00000(4) | 0.00001(6) | 0.00001(5) |

|      | **ZnSb 225K**   | **ZnSb 300K**    | **ZnSb 300K**   | **ZnSb 350K**   | **ZnSb 400K**   |
|------|-----------------|------------------|-----------------|-----------------|-----------------|
| atom | Sb              | Sb               | Sb              | Sb              | Sb              |
| $x$  | 0.14307(3)      | 0.14307(2)       | 0.14310(3)      | 0.14309(3)      | 0.14315(3)      |
| $y$  | 0.08214(2)      | 0.08219(2)       | 0.08218(2)      | 0.08222(2)      | 0.08227(2)      |
| $z$  | 0.10831(2)      | 0.10828(2)       | 0.10832(2)      | 0.10831(3)      | 0.10827(3)      |
| $U_{eq}$  | 0.00868(2)      | 0.011174(17)     | 0.01099(3)      | 0.01296(3)      | 0.01488(3)      |
| $U_{11}$ | 0.00798(4)      | 0.01032(3)       | 0.00992(5)      | 0.01164(4)      | 0.01343(4)      |
| $U_{22}$ | 0.00889(4)      | 0.01149(3)       | 0.01148(5)      | 0.01344(4)      | 0.01540(5)      |
| $U_{33}$ | 0.00916(4)      | 0.01171(3)       | 0.01158(5)      | 0.01380(4)      | 0.01582(5)      |
| $U_{12}$ | 0.000272(18)    | 0.000332(16)     | 0.00033(3)      | 0.00038(3)      | 0.00042(3)      |
| $U_{13}$ | 0.000195(18)    | 0.000287(15)     | 0.00024(3)      | 0.00034(3)      | 0.00039(3)      |
| $U_{23}$ | 0.000119(18)    | 0.000155(16)     | 0.00018(2)      | 0.00019(3)      | 0.00018(3)      |
| $C_{111}$ | 0.00000(6)     | -0.00002(4)      | -0.00004(8)     | -0.00004(8)     | 0.00006(8)      |
| $C_{112}$ | 0.00003(3)     | 0.00005(2)       | 0.00002(4)      | 0.00009(4)      | 0.00013(4)      |
| $C_{113}$ | 0.00003(3)     | 0.000046(19)     | 0.00012(4)      | 0.00014(3)      | 0.00009(4)      |
| $C_{122}$ | 0.00002(2)     | 0.000049(17)     | 0.00009(3)      | 0.00006(3)      | 0.00009(3)      |
| $C_{123}$ | 0.000005(14)   | -0.000001(11)    | 0.00001(2)      | -0.00001(2)     | 0.00002(2)      |
| $C_{133}$ | 0.00003(2)     | 0.000010(14)     | 0.00004(3)      | 0.00005(3)      | 0.00005(3)      |
| $C_{222}$ | 0.00008(3)     | 0.00008(3)       | 0.00003(4)      | 0.00004(4)      | 0.00009(5)      |
| $C_{223}$ | 0.000013(18)   | 0.000004(14)     | 0.00003(2)      | 0.00004(2)      | 0.00003(3)      |
| $C_{233}$ | 0.000010(18)   | 0.000039(13)     | 0.00004(2)      | 0.00008(2)      | 0.00010(2)      |
| $C_{333}$ | -0.00001(3)    | -0.00003(2)      | -0.00002(4)     | 0.00000(4)      | -0.00004(4)     |
| atom | Zn              | Zn               | Zn              | Zn              | Zn              |
| $x$  | 0.54367(6)      | 0.54444(6)       | 0.54454(7)      | 0.54516(7)      | 0.54571(8)      |
| $y$  | 0.61041(4)      | 0.61099(4)       | 0.61092(5)      | 0.61129(5)      | 0.61170(5)      |
| $z$  | 0.63007(4)      | 0.63044(4)       | 0.63034(4)      | 0.63058(5)      | 0.63077(5)      |
| $U_{eq}$ | 0.01375(4)      | 0.01827(3)       | 0.01808(5)      | 0.02159(5)      | 0.02514(6)      |
| $U_{11}$ | 0.01556(7)      | 0.02100(6)       | 0.02054(9)      | 0.02448(10)     | 0.02869(11)     |
| $U_{22}$ | 0.01273(6)      | 0.01682(6)       | 0.01687(8)      | 0.02007(9)      | 0.02322(10)     |
| $U_{33}$ | 0.01297(7)      | 0.01700(6)       | 0.01682(8)      | 0.02023(9)      | 0.02350(10)     |
| $U_{12}$ | 0.00232(5)      | 0.00341(4)       | 0.00339(6)      | 0.00418(7)      | 0.00506(8)      |
| $U_{13}$ | 0.00145(5)      | 0.00216(4)       | 0.00210(6)      | 0.00260(7)      | 0.00321(8)      |
| $U_{23}$ | 0.00167(4)      | 0.00234(4)       | 0.00231(6)      | 0.00290(6)      | 0.00352(7)      |
| $C_{111}$ | 0.00130(15)    | 0.00211(13)      | 0.0024(2)       | 0.0034(3)       | 0.0044(3)       |
| $C_{112}$ | 0.00055(7)     | 0.00112(6)       | 0.00108(10)     | 0.00167(11)     | 0.00215(12)     |
| $C_{113}$ | 0.00049(6)     | 0.00101(5)       | 0.00093(9)      | 0.00153(10)     | 0.00190(11)     |
| $C_{122}$ | 0.00052(6)     | 0.00077(5)       | 0.00080(8)      | 0.00109(9)      | 0.00152(10)     |
| $C_{123}$ | 0.00017(3)     | 0.00034(3)       | 0.00034(5)      | 0.00046(6)      | 0.00065(6)      |
| $C_{133}$ | -0.00012(5)    | -0.00017(4)      | -0.00006(8)     | -0.00014(8)     | -0.00018(9)     |
| $C_{222}$ | 0.00001(8)     | 0.00002(7)       | 0.00003(10)     | -0.00004(11)    | 0.00012(12)     |
| $C_{223}$ | 0.00008(4)     | 0.00020(4)       | 0.00018(5)      | 0.00021(6)      | 0.00019(7)      |
| $C_{233}$ | 0.00013(4)     | 0.00029(3)       | 0.00030(5)      | 0.00036(6)      | 0.00053(7)      |
| $C_{333}$ | 0.00004(6)     | 0.00011(5)       | 0.00011(8)      | 0.00012(9)      | 0.00024(10)     |

**S5: Plot of temperature dependent anharmonic ADP parameters**

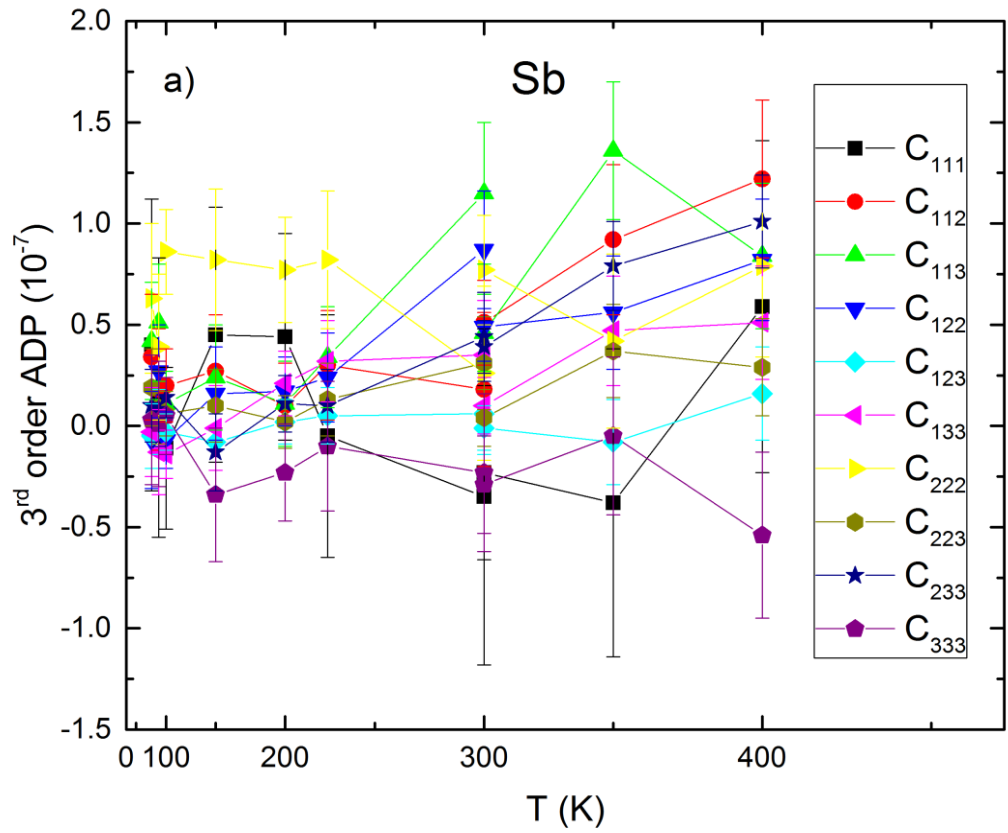

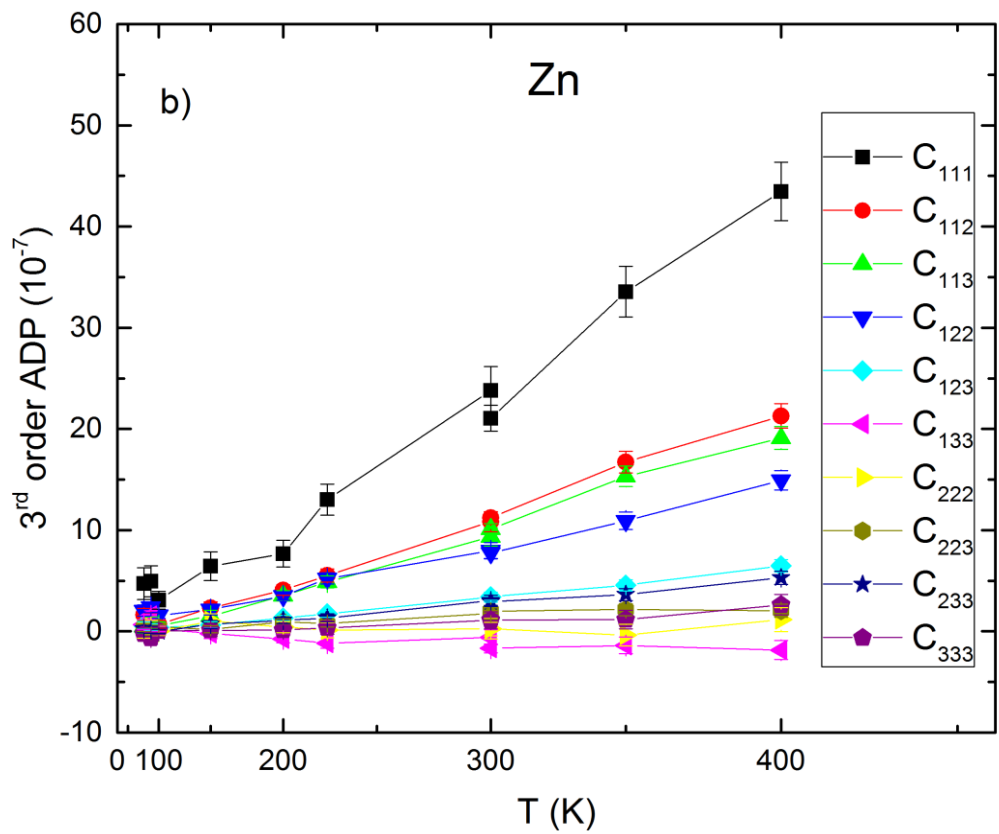

**S6: Plot of temperature dependent off-diagonal harmonic ADP parameters**

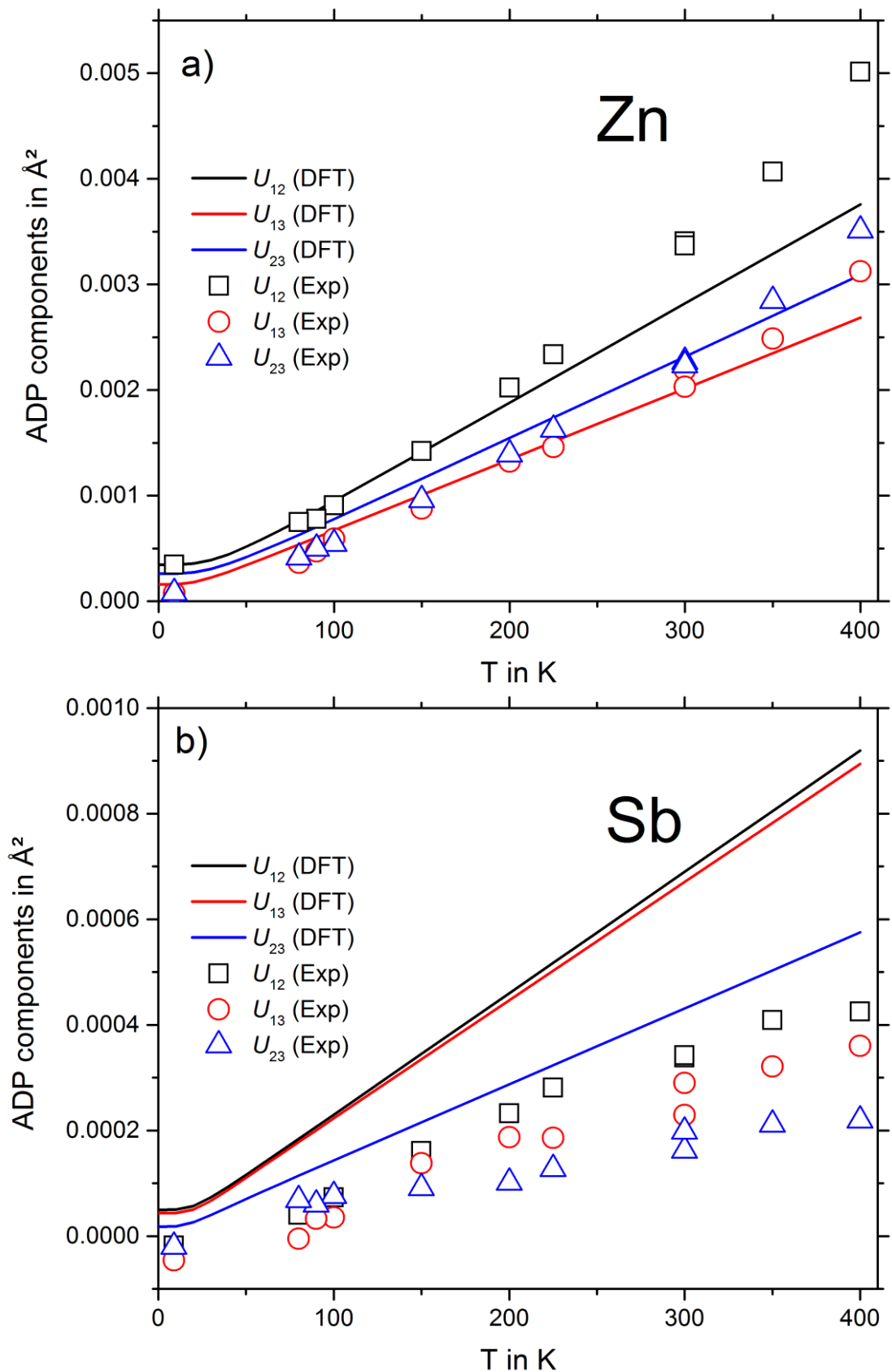